\documentclass[a4paper,11pt,superscriptaddress,preprintnumbers,floatfix]{article}
\pdfoutput=1 
\usepackage{jheppub} 
\usepackage{bm}
\usepackage{qcircuit}
\usepackage{subcaption}
\usepackage{overpic}
\usepackage[T1]{fontenc}
\usepackage{epsfig}
\usepackage{graphicx}
\usepackage[english]{babel}
\usepackage{hyphenat}
\usepackage{amsmath}
\usepackage{amssymb}
\usepackage{mathtools}
\usepackage{mathrsfs}
\usepackage{slashed}
\usepackage{epstopdf}
\usepackage{xcolor}
\usepackage{braket}
\usepackage{booktabs}
\definecolor{lcolor}{rgb}{0.,0.0,0.}
\definecolor{citcolor}{rgb}{0,0.,0.5}
\usepackage{multirow}
\usepackage{ltablex}
\usepackage{soul}
\usepackage{cleveref}
\crefname{section}{section}{sections}
\crefname{figure}{figure}{figures}
\crefname{appendix}{Appendix}{Appendices}
\crefname{equation}{eq.}{eqs.}
\crefname{table}{Table}{Tables}
\Crefname{section}{Section}{Sections}
\Crefname{figure}{Figure}{Figures}
\Crefname{appendix}{Appendix}{Appendices}
\Crefname{equation}{Equation}{Equations}
\Crefname{table}{Table}{Tables}

\usepackage[e]{esvect}

\def\cG{{\cal G}}
\def\cA{{\cal A}}
\def\cC{{\cal C}}

\def\cP{{\cal P}}
\def\cK{{\cal K}}
\def\cM{{\cal M}}
\def\cD{{\cal D}}

\def\cQ{{\cal Q}}

\newcommand{\secn}[1]{Section~1}
\newcommand{\appn}[1]{Appendix~1}

\long\def\comment#1{ }

\def\Tr{\text{Tr}}

\def\and{\quad\text{and}\quad}

\def\BK{{\boldsymbol K}}

\def\q{{\boldsymbol q}}
\def\0{{\boldsymbol 0}}
\def\1{{\boldsymbol 1}}
\def\p{{\boldsymbol p}}

\def\k{{\boldsymbol k}}
\def\x{{\boldsymbol x}}
\def\y{{\boldsymbol y}}

\def\r{{\boldsymbol r}}

\def\u{{\boldsymbol u}}
\def\v{{\boldsymbol v}}

\def\0{{\boldsymbol 0}}

\def\P{{\boldsymbol P}}

\def\bkappa{{\boldsymbol \kappa}}

\newcommand{\expconfig}[1]{{\langle #1 \rangle}}

\newcommand{\occf}{{\omega}}

\newcommand{\ind}{{\beta}} 
\newcommand{\pd}{{\vphantom{\dagger}}}
\newcommand*\diff{\mathop{}\!\mathrm{d}}

\renewcommand\u{\upsilon}

\newcommand\G{\Gamma}

\def\u{{\boldsymbol u}}

\renewcommand{\part}{{\rm part}}

\newcommand{\be}{\begin{equation}}
\newcommand{\ee}{\end{equation}}
\newcommand{\bes}{\begin{subequations}}
\newcommand{\ees}{\end{subequations}}
\newcommand{\bea}{\begin{eqnarray}}
\newcommand{\eea}{\end{eqnarray}}

\newcommand{\nn}{\nonumber \\}

\def\bea#1\eea{\begin{align}#1\end{align}}
\newcommand{\bef}{\begin{figure}[h!tb]\centering}
\newcommand{\eef}{\end{figure}}

\newcommand{\GeV}{{{\,}\textrm{GeV}}}
\allowdisplaybreaks

\title{Quantum simulating multi-particle processes in high energy nuclear physics: dijet production and color (de)coherence} 

\author[a,*]{Jo\~{a}o Barata,\note[*]{Authors are listed in alphabetical order.}}
\author[b,*]{Meijian Li,}
\author[c,b,*]{Wenyang Qian,}
\author[b,d,*]{Carlos A. Salgado,}
\author[e,f,g,*]{Jo\~{a}o M. Silva}

\affiliation[a]{CERN, Theoretical Physics Department, CH-1211 Geneva 23, Switzerland}
\affiliation[b]{Instituto Galego de F{\'{i}}sica de Altas Enerx{\'{i}}as,  Universidade de Santiago de Compostela, Santiago de Compostela 15782, Galicia, Spain}
\affiliation[c]{Institute of Particle Physics and Key Laboratory of Quark and Lepton Physics (MOE),
Central China Normal University, Wuhan, 430079, Hubei, China}
\affiliation[d]{Axencia Galega de Innovacion (GAIN), Xunta de Galicia, Galicia, Spain}
\affiliation[e]{Laboratório de Instrumentação e Física Experimental de Partículas (LIP), Av. Prof. Gama Pinto, 2, 1649-003 Lisbon, Portugal}
\affiliation[f]{Departamento de Física, Instituto Superior Técnico (IST), Universidade de Lisboa, Av. Rovisco Pais 1, 1049-001 Lisbon, Portugal}
\affiliation[g]{Departamento de Física Teórica y del Cosmos, Universidad de Granada, Campus de Fuentenueva, E-18071 Granada, Spain}

\emailAdd{joao.lourenco.henriques.barata@cern.ch}
\emailAdd{meijian.li@usc.es}
\emailAdd{wqian@ccnu.edu.cn}
\emailAdd{carlos.salgado@usc.es}
\emailAdd{joao.m.da.silva@tecnico.ulisboa.pt}

\preprint{CERN-TH-2026-082}

\abstract{
Hard scattering events in high-energy collisions produce highly virtual partons that subsequently fragment into collimated hadronic cascades. When such partonic showers evolve in a QCD medium, as in deep-inelastic scattering or heavy-ion collisions, the resulting multi-particle distributions encode information about the surrounding matter. Decades of theoretical developments have led to a consistent and order-by-order improvable perturbative description of the shower. This description needs, however, the non-perturbative input that encodes the structure of the hadronic  matter. The determination of such input remains challenging within conventional computational approaches, thereby limiting the  applicability of the approach. In this work, we develop a framework that employs quantum simulation techniques to compute multi-particle processes in such environments by mapping partonic cross-sections to quantum circuits. As benchmarks, we analyze dipole formation and the QCD antenna radiation pattern at leading order in the strong coupling constant, comparing the results with analytic estimates in simplified limits. The quantum circuit formulation here introduced naturally extends to higher perturbative orders and enables amplitude-level computations in complex matter backgrounds. This provides a systematic foundation for applying quantum information science methods to study multi-particle dynamics in QCD media.}
\begin{document}

\maketitle

\section{Introduction}\label{sec:intro}
Hard processes in high-energy nuclear collisions provide a direct window into the short-distance structure of nuclear matter, much as Rutherford scattering once revealed the internal composition of the atom. A paradigmatic example is deep-inelastic scattering (DIS), in which a high-energy electron transfers a large momentum to a proton or nuclear target, thereby resolving its subnuclear constituents and probing the microscopic dynamics of quantum chromodynamics (QCD)~\cite{Morreale:2021pnn}. Closely related information is obtained from the production of energetic hadrons and jets in nuclear environments, such as the quark-gluon plasma (QGP) generated in heavy-ion collisions, where interactions between the emerging partonic cascade and the surrounding medium are imprinted on the final-state particle distributions~\cite{Busza:2018rrf}. Experimentally, the study of such processes is a major focus of the physics programs at Relativistic Heavy Ion Collider (RHIC) and the Large Hadron Collider (LHC)~\cite{Arslandok:2023utm}, and will remain a central component of the future Electron Ion Collider (EIC) program~\cite{Accardi:2012qut} and the Electron Ion Collider in China (EicC)~\cite{Anderle:2021wcy}.

An important theoretical challenge in describing such processes lies in the treatment of the non-perturbative dynamics governing the interaction between energetic probes and the nuclear medium. In the high-energy limit, a useful organizing picture follows Feynman's parton-model intuition, in which the interaction is dominated by soft gluonic modes sourced by the hard constituents of the bulk matter. In the context of DIS at small Bjorken-$x$, this scale separation is implemented in the Color Glass Condensate (CGC) framework~\cite{McLerran:1994vd,McLerran:1993ka,McLerran:1993ni}, with closely related formulations applying to QCD jets propagating through the QGP produced in heavy-ion collisions~\cite{Mehtar-Tani:2013pia,Blaizot:2012fh}. Within this picture, perturbatively calculable hard contributions factorize from the non-perturbative structures encoding the interaction with the nuclear target, which are expressed in terms of gauge-invariant correlation functions of non-local operators tracking the space-time and color evolution of energetic charges traversing the bulk medium~\cite{Gelis:2010nm}.  
Despite substantial progress, see e.g.~\cite{Kuzmin:2025fyu,Barata:2021wuf,Mehtar-Tani:2019tvy,Andres:2020vxs,Alba:future, Andres:2026qrt,Caron-Huot:2010qjx,Isaksen:2023nlr,Isaksen:2020npj,Kuzmin:2024smy,Andres:2025prc,Apolinario:2014csa,Feal:2018sml,Sadofyev:2021ohn,Sievert:2018imd,Sievert:2019cwq,Altinoluk:2021lvu}, the calculation of these non-perturbative objects remains challenging, typically involving additional simplifying assumptions or approximations, see e.g.~\cite{Apolinario:2014csa,Feal:2018sml,Isaksen:2023nlr,Caron-Huot:2010qjx,Mehtar-Tani:2019tvy,Barata:2020sav,Adhya:2024nwx,Barata:2025agq, Barata:2024bqp,Barata:2023qds}. 
Achieving systematic improvements in the computational precision of these correlations and designing strategies to treat general multi-particle amplitudes in realistic matter backgrounds is of crucial phenomenological relevance.

One approach to study these high-energy nuclear processes, complementary to more traditional methods found in the literature~\cite{Apolinario:2014csa,Feal:2018sml,Isaksen:2023nlr,Caron-Huot:2010qjx,Mehtar-Tani:2019tvy}, is to directly compute them using fully non-perturbative methods, based on the light-front Hamiltonian formalism~\cite{Brodsky:1997de}. This approach has been recently applied to study the real-time evolution of QCD jets in a color medium~\cite{Li:2020uhl,Li:2021zaw,Li:2023jeh,Li:2025wzq}, going beyond the leading eikonal order and avoiding perturbative expansions.
This framework has been further extended using quantum computing techniques~\cite{Barata:2021yri,Barata:2022wim,Barata:2023clv,Qian:2024gph,Wu:2024adk, Castro:2025ocx}, opening new avenues for treating the full complexity of the problem on quantum hardware.\footnote{See also e.g.~\cite{Barata:2025jhd,Rodrigo:2026xom,Clemente:2022nll,Barata:2024apg,Barata:2025hgx,Li:2025sgo,Du:2023ewh,Ikeda:2024rzv,Su:2024uuc,Chen:2024pee,Shao:2025obi,Galvez-Viruet:2025rmy,Galvez-Viruet:2025ket,Qian:2021jxp,Li:2022lyt,Li:2023kex,Li:2021kcs,Li:2024nod,Kreshchuk:2020dla,Kreshchuk:2020aiq,Du:2020glq,Qian:2024xnr,Grieninger:2024cdl,Banuls:2025wiq,Chai:2023qpq,Barata:2025rjb,Grieninger:2025mbm} for related discussion.}
While these developments represent important progress, the full simulation of jet fragmentation is computationally hard to achieve in current quantum platforms. 
Nonetheless, these novel methods can instead be applied to systematically evaluate specific multi-particle correlators, with perturbatively tractable contributions handled separately using conventional methods, thereby enabling a more targeted and computationally feasible approach to the problem.

In this work, we illustrate this strategy by introducing a framework in which the non-perturbative correlation functions encoding the properties of the medium are evaluated directly in the Hamiltonian formalism at the amplitude level, in close connection with the light-front QCD Hamiltonian approach~\cite{Li:2020uhl,Li:2021zaw,Li:2023jeh,Li:2025wzq} and the quantum simulation framework of~\cite{Qian:2024gph}.
Rather than working directly with ensemble-averaged quantities, as is commonly done in the literature, we propagate quantum states in color and space-time and extract the relevant correlation functions from the corresponding measured states, prepared in a quantum circuit. This formulation automatically resolves the full color structure of multi-particle processes and accommodates arbitrary statistical properties of the soft gluon background.
Because the evolution is implemented as a unitary real-time process, the method is naturally aligned with quantum simulation in the Hamiltonian picture and is therefore well suited for implementation on quantum hardware. 

This work is organized as follows. In \cref{sec:mapping_to_QIS} we discuss the general mapping between high-energy multi-particle production and quantum-circuit-based simulation. We then construct explicit realizations for leading-order dijet production\footnote{We consider the leading-order process $\gamma \to q\bar q$ in the collinear limit, corresponding to the production of a highly boosted color dipole. Although this process is often referred to as "dijet production" in the DIS literature~\cite{Dominguez:2011wm}, where two jets can naturally be distinguished at forward rapidity, in the remaining of this work we will rather use the term "dipole formation". The main purpose is to also make the connection with heavy-ion collisions, where the term "dijet production" in the collinear limit becomes less adequate.} from a virtual photon, for which the calculation can be benchmarked at leading eikonal accuracy, and for the color (de)coherence of QCD antennas in a dense medium. Details of the mapping to a qubit-based system are presented in \cref{sec:method}, while numerical results and comparisons with analytic expectations are given in \cref{sec:results}. We summarize our findings and discuss possible future applications in \cref{sec:conclusion}.

\begin{figure}[htp!]
    \centering
    \begin{overpic}[width=\textwidth]{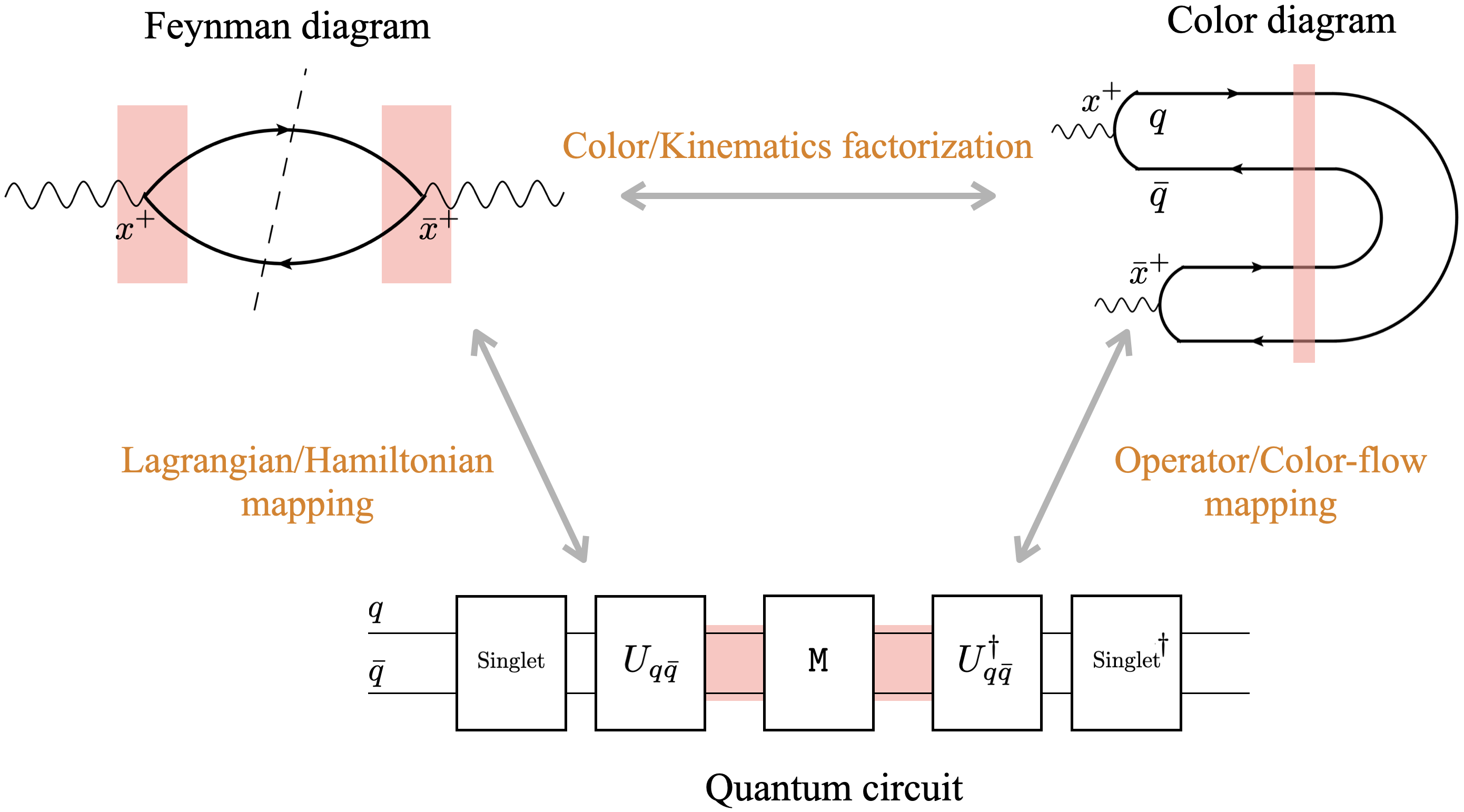}
        \put(5.6,53){${(a)}$}
        \put(75.8,53.5){${(b)}$}
        \put(44.3,1.2){${(c)}$}
    \end{overpic}
    \caption{\textbf{Mapping structure for dipole formtion at leading order.} The only relevant Feynman diagram $(a)$ at leading order includes the contribution where the outgoing quarks branch in the presence of a background field (pink box), which can, e.g., represent either the color field associated to the proton in DIS or the QGP produced in heavy ion collisions. Due to color precession, the dynamics can be conveniently encapsulated in a diagrammatic representation of the color flow $(b)$, aligned along the contour. This form makes more evident the \textit{factorization} between the initial pair production and the subsequent evolution in the presence of the bulk. Finally, both these formulations can be mapped to a quantum circuit, where the interaction with the bulk is introduced via classical communication channels $(c)$. 
    }
    \label{fig:dijet} 
\end{figure}

\section{High energy partonic processes in nuclear matter}\label{sec:mapping_to_QIS}

In this section we lay out the connection between high-energy scattering processes involving multi-particle final states and quantum simulation. 
In particular, we are interested in the regimes where the fragmentation process takes place in the presence of a dense colored background field, which can be treated in classical Yang-Mills theory, as commonly found in e.g., small-$x$~\cite{Gelis:2010nm,Morreale:2021pnn}, and jet quenching literature~\cite{Casalderrey-Solana:2007knd,Mehtar-Tani:2013pia}. We explain the mapping via two example processes. The overall structure is summarized in \cref{fig:dijet}.

\subsection{Collinear dipole formation in nuclear matter}\label{subsec:dijet}

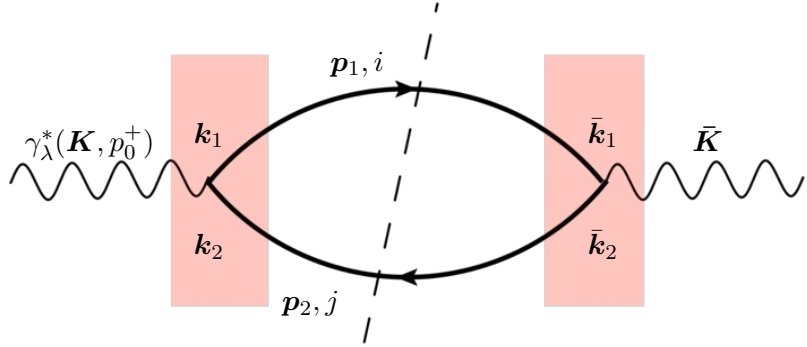
\begin{figure}[t!]
    \centering
    \begin{overpic}[width=0.7\textwidth]{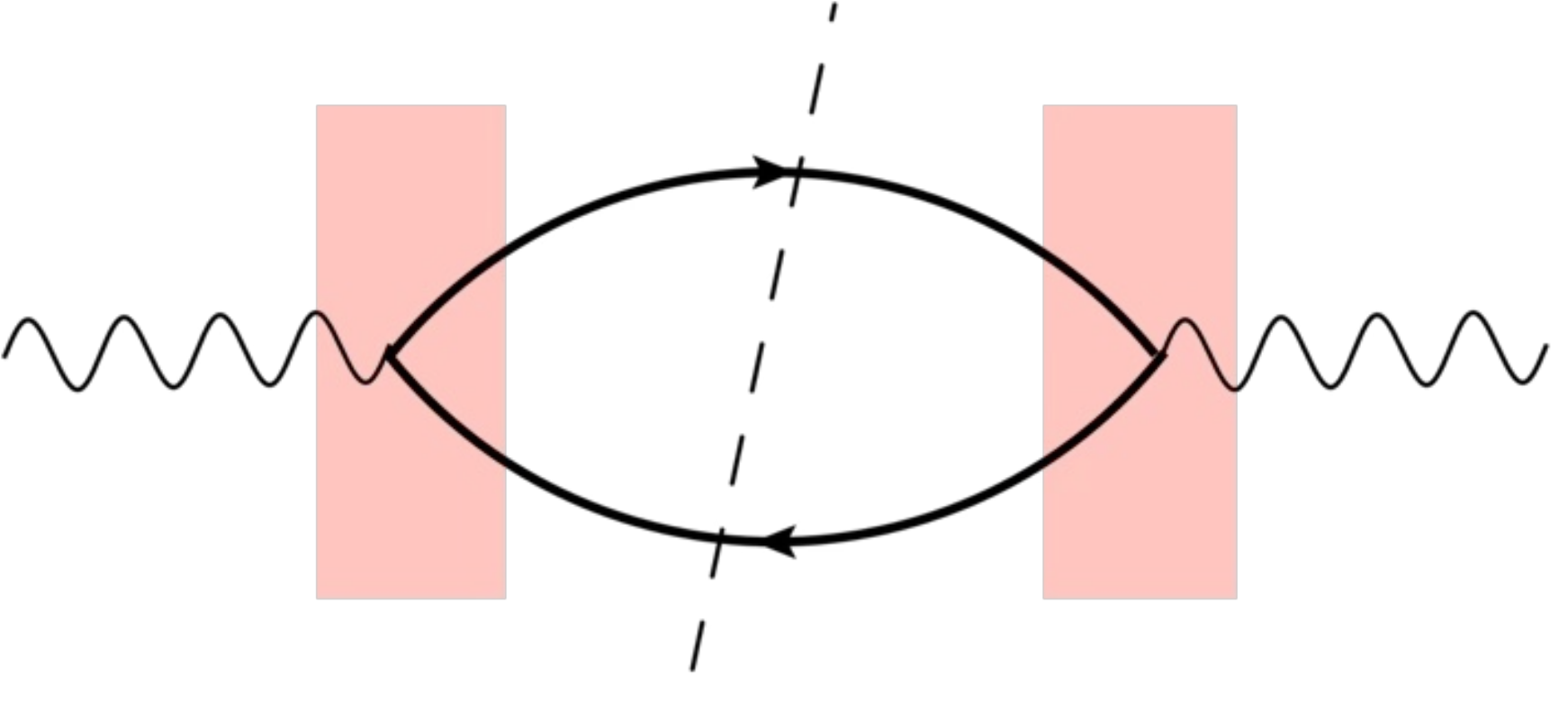}
        \put(2,26){$\gamma^*_\lambda (\boldsymbol{K}, p_0^+)$}
        \put(23,27){$\k_1$}
        \put(23,13){$\k_2$}
        \put(40,36){$\p_1, i$}
        \put(34,6){$\p_2, j$}
        \put(72,27){$\bar\k_1$}
        \put(72,13){$\bar\k_2$}
        \put(85,26){$\boldsymbol{\bar K}$}
    \end{overpic}
\caption{
\textbf{Leading order diagram for dipole formation in the presence of a nuclear target.} 
The dashed line separates the amplitude from the conjugate amplitude. 
Particle momenta and color indices are labeled, and the background field is
indicated by the pink box.
}
\label{fig:Feynman_qqbar}
\end{figure}

To explain the mapping between the QCD picture and one of quantum circuits, we consider the simple process of dipole formation from a virtual photon in the presence of a nuclear medium, at leading order in the strong coupling constant, as illustrated in \cref{fig:dijet} $(a)$ and in \cref{fig:Feynman_qqbar}. As mentioned above, this is the leading contribution for this process both in the context of deep-inelastic scattering~\cite{Dominguez:2011wm,Caucal:2021ent} and in proton-ion and heavy ion collisions~\cite{Mehtar-Tani:2010ebp,Armesto:2011ir,Dominguez:2019ges}.\footnote{
We note that although the computations in these different contexts start from slightly different approximations, leading to important differences in the final expressions, these distinctions are not relevant for the quantum circuit implementation.
}

At leading eikonal accuracy, the amplitude for this process reads, see e.g.~\cite{Isaksen:2023nlr,Dominguez:2019ges}:
\begin{multline}\label{eq:gamma_qqbar_amplitude}
    i\mathcal{M}^{ij,hh'}(\p_1, \p_2, z, p^+_0) = \frac{e^{i(\frac{\p_1^2}{2p_1^+}+\frac{\p_2^2}{2p_2^+})x^+_{\infty}}}{2p_0^+}\sum_{\lambda=\pm 1}\int_{x^+,\k_1,\k_2}e^{-\epsilon x^+}\,ie V^{\lambda,h,h'}(\bkappa,z)
    \\
      \times 
      \mathcal{M}^{\lambda}_0(\boldsymbol{K},p_0^+)
  e^{-i\frac{\boldsymbol{K}^2}{2p_0^+}x^+} G^{ik}(x^+_{\infty},\p_1;x^+,\k_1;p_1^+)\bar G^{kj}(x^+_{\infty},\p_2;x^+,\k_2;p_2^+)\;,
\end{multline}
where the splitting time $x^+\in (0,x^+_{\infty})$ and one should take $x^+_{\infty} \rightarrow +\infty$ and $\epsilon \rightarrow 0$ after carrying out all integrals according to the appropriate adiabatic prescription~\cite{Wiedemann:2000za}.
We use the notations $\int_\q = \int \diff^2\q (2\pi)^{-2}$, $\int_\x = \int \diff^2\x $ for transverse integrals, and $\int_{x^+} = \int \diff x^+$ accordingly for the time integrations. Here $\boldsymbol{K} = \k_1+\k_2$ is the photon's transverse momentum, $p_0^+$ its longitudinal momentum,  and $\lambda=\pm 1$ its transverse polarization, i.e., we assume a nearly on-shell photon; $\k_1 $($\k_2$) is the transverse momentum of the (anti)quark at $x^+$, $\p_1$ ($\p_2$) its transverse momentum at $x^+_\infty$, $p^+_1$($p^+_2$) its longitudinal momentum, $h(h')=\pm1$ its helicity state, and $i$($j$) its final color index, as in \cref{fig:Feynman_qqbar}. 
Note that the longitudinal momentum does not change via interaction with the background field, such that $p_0^+=p^+_1+p^+_2$.
For convenience, we define the energy fraction and relative momentum
\begin{align}
 z = p_1^+/p_0^+\,,
 \qquad \bkappa = (1-z)\k_1-z\k_2\;.
\end{align}
The photon-to-dipole vertex, for a nearly on-shell photon, i.e., a highly boosted off-shell  photon $p_0^2/(p_0^+)^2 \ll 1$, is given by 
\begin{align}
\begin{split}
    & V^{\lambda,h,h'}(z,\bkappa) =\frac{2\delta^{h,-h'}}{\sqrt{z(1-z)}}
    \left(
    z\delta^{\lambda, h}-(1-z)\delta^{\lambda, -h}
    \right)(\bkappa\cdot\boldsymbol{\epsilon}^{\lambda})\;.
\end{split}
\end{align}
where $\bm\epsilon^\lambda=(1, i\lambda)/\sqrt{2}$.
Then, $\mathcal{M}^{\lambda}_0(\vec{p}_0) = \mathcal{M}^{\mu}_0(\vec p_0)\epsilon_{\mu}^{*,\lambda}(\vec p_0)$ with $\vec p_0 = (\BK,p_0^+)$
is the initial amplitude producing the photon at time $x^+=0$, and $e^{-i\boldsymbol{K}^2/(2p_0^+)x^+}$ comes from the free propagator of the photon.
Finally, the Green's function $G^{ik}$ ($\bar G^{kj}$) describes the propagation of the (anti)quark in the presence of the background at fixed light-cone energy.
At leading eikonal order, i.e., leading order in $1/p^+$, this object reduces to an effective quantum mechanical propagator, which in the light-cone gauge and in coordinate space takes the form~\cite{Casalderrey-Solana:2007knd}
\begin{align}\label{eq:def_in_medium_propagator}
	\cG^{ij}(x_2^+,\x_2;x_1^+,\x_1|p^+) = \int^{\boldsymbol{r}(x_2^+) = \boldsymbol{x_2}}_{\boldsymbol{r}(x_1^+)= \boldsymbol{x_1}}\mathcal{D}\boldsymbol{r}\exp\Big\{i\frac{p^+}{2}\int_{x_1^+}^{x_2^+}\diff s^+ \, \dot{\boldsymbol{r}}^2\Big\}U^{ij}(x_1^+,x_2^+,\boldsymbol{r}) \, ,
\end{align}
with $\dot{\boldsymbol{r}}\equiv d{\boldsymbol{r}}/ds^+$, and the endpoints denoting the starting and final transverse positions of the particle's evolution in light-cone time $x^+$ at fixed light-cone energy $p^+$. The path integral accounts for the accumulation of transverse momentum by interaction of the quark with the surrounding matter. Furthermore, its color state is continuously rotated and this is encapsulated by the path-ordered Wilson line
\begin{equation}\label{eq:def_wilson_line}
    U^{ij}(x^+,y^+,\boldsymbol{r}) = \mathcal{P}\exp\Big\{ig\int_{x^+}^{y^+}\diff s^+\, \cA_a^{-}(s^+, \boldsymbol{r}(s^+))t^a_{ij}\Big\}\, .
\end{equation}
The anti-quark propagator is given by $\bar \cG$, where one replaces $U \rightarrow U^\dagger$ in the expression for $\cG$. The propagators in \cref{eq:gamma_qqbar_amplitude} are given by a Fourier transform to momentum space of the propagator in \cref{eq:def_in_medium_propagator}
\begin{align}
    G^{ij}(x_2^+,\p_2;x_1^+,\p_1|p^+) = \int_{\x_1, \x_2} e^{i\x_1\cdot\p_1}e^{-i\x_2\cdot\p_2}\cG^{ij}(x_2^+,\x_2;x_1^+,\x_1|p^+) \, .
\end{align}
In the vacuum limit this propagator reduces to
\begin{align}
    G_0^{ij}(x_2^+,\p_2;x_1^+,\p_1|p^+) = \delta^{ij}(2\pi)^2\delta^2(\p_2-\p_1)e^{-i\frac{\p_1^2}{2p^+}(x_2^+-x_1^+)} \, .
\end{align}

As mentioned above, the nuclear target is treated in the limit of a large gluon occupation number, such that it admits a semi-classical treatment. To that effect, one expands the gauge field around the classical solution $\cA$, 
induced by a stochastic distribution of color sources. As such, the computation of any in-medium process should involve an average over possible configurations of the color sources generating this field. We assume that $\mathcal{A}$ exists during the light-cone time interval $x^+\in (0,L)$. Thus, by splitting the $x^+$ integration into two regions $x^+\in (0,L)$ and $x^+\in(L,x^+_{\infty})$, the amplitude can be expressed as a summation of two contributions, $\cM = \cM^{\rm in}+\cM^{\rm out}$,  where the $\gamma\rightarrow q\bar q$ splitting occurs either inside or outside the finite-size medium.
Upon squaring the amplitude, one has three contributions, 
\begin{align}
\cM\cM^{\dagger} = \cM^{\rm in}\cM^{\rm in,\dagger}+2{\rm Re}(\cM^{\rm in}\cM^{\rm out,\dagger})+\cM^{\rm out}\cM^{\rm out,\dagger}\;,
\end{align}
each giving rise to different multi-point correlator. 
The differential cross-section reads:
\begin{align}\label{eq:gamma_qqbar_fullxsec}
\begin{split}
 \frac{ (2\pi)^3 \diff \sigma}{\diff^3\Omega_0\,\diff z\,\diff^2\p} 
    & =    
    \frac{1}{ 2z(1-z)}
   \sum_{i,j}^{N_c}\sum_{h,h'=\pm1} \lvert\cM^{ij,hh'}(\p_1, \p_2, z, p^+_0)\rvert^2\\
    & =  N_c e^2 P_{q\gamma}(z)\sum_{\lambda\pm1}\Bigg(\frac{1}{2\omega^2}{\rm Re}\int_0^L \diff x^+\int_t^L \diff\bar x^+\int_{\k_1,\k_2,\bar \k_1,\bar \k_2} 
    \\
     &
     \mathcal{M}_0^\lambda(\BK)\mathcal{M}_0^{\dagger,\lambda}(\bar \BK)
     (\bkappa \cdot \bar{\bkappa})e^{-i\frac{\BK^2}{2p_0^+}x^+}e^{i\frac{\bar\BK^2}{2p_0^+}\bar x^+}\mathcal{C}_4 
     -\frac{1}{\p^2\omega}{\rm Re}\,i\int_0^L \diff x^+\int_{\k_1,\k_2}
     \\
     &
     \mathcal{M}_0^\lambda(\BK)\mathcal{M}_0^{\dagger,\lambda}(\P)(\bkappa \cdot \p)e^{-i\frac{\BK^2}{2p_0^+}x^+}e^{i\frac{\P^2}{2p_0^+}L}\mathcal{C}_2 +\frac{|\cM_0^\lambda(\P)|^2}{\p^2}\Bigg)\;.
     \end{split}
\end{align}
Here, we use an overbar to denote momentum variables $\bar \k_1$ and $\bar \k_2$ in the complex conjugate amplitude. Accordingly, $\bar\bkappa = (1-z)\bar\k_1 - z\bar\k_2$, $\bar\BK = \bar\k_1 + \bar\k_2$ and
we used $\p = (1-z)\p_1 - z\p_2$, $\P = \p_1 + \p_2$. 
The initial measure reads $\diff^3\Omega_0 =\diff p_0^+\diff^2\P/(2 p_0^+(2\pi)^3)$ and we defined $\omega\equiv z(1-z)p_0^+$. 
Note that in the squared vertex only terms where the photon polarisation $\lambda$ is the same in the amplitude and complex conjugate survive (i.e. no spin correlations, see e.g.~\cite{Silva:2025dan}), as well as scalar momenta dependencies of the type $\bkappa\cdot\bar\bkappa$, $\bkappa\cdot\p$ and $\p^2$. In order for this to hold, we are implicitly assuming an azimuthal integration at the end of the calculation, as well as a medium that respects rotation invariance and parity-evenness (invariance under $x(y)\rightarrow -x(-y)$). 
We have further introduced the Altarelli-Parisi splitting function for the process $\gamma\rightarrow q\bar q$
\begin{align}
    P_{q\gamma}(z)=z^2+(1-z)^2\;.
\end{align}
Note that the last term in \cref{eq:gamma_qqbar_fullxsec} is exactly the pure vacuum contribution, corresponding to the splitting $\gamma\rightarrow q\bar q$ in the absence of a medium. In \cref{eq:gamma_qqbar_fullxsec}, the production of the initial antenna is described perturbatively, while the propagation in the matter is encapsulated into the multi-point correlators $\mathcal{C}_4$ and $\mathcal{C}_2$. 
These correlators are non-perturbative and computing them is equivalent to evaluating the full cross-section at this order:
\begin{align}\label{eq:C4:C2:1}
\begin{split}
    \cC_4 =  & \frac{1}{N_c}\Big\langle \Tr \left(\bar G^{\dagger}(L,\p_2;\bar x^+,\bar\k_2; p_2^+)G^{\dagger}(L,\p_1;\bar x^+,\bar\k_1;p_1^+)G(L,\p_1;x^+,\k_1;p_1^+)
    \right.\\
    &\left.
    \bar G(L,\p_2;x^+,\k_2;p_2^+)\right) \Big\rangle\,,\\ 
   \cC_2 =  & \frac{1}{N_c}\Big\langle \Tr \left(G(L,\p_1;x^+,\k_1; p_1^+)\bar G(L,\p_2;x^+,\k_2;p_2^+)\right)\Big\rangle\;,
\end{split}
\end{align}
where $\langle \,\cdots \rangle$ denotes the stochastic averaging over the configurations of the field $\mathcal{A}$.

We now connect this standard perturbative discussion to a language based on quantum circuits. First, we find it convenient to relate the description in terms of Feynman diagrams to that of a color diagram, see  \cref{fig:dijet} $(b)$, as introduced in e.g.,~\cite{Dominguez:2011wm}. Here one highlights the evolution in color space, while the evolution in space is suppressed. 
It is clear that the cross-section involves the evolution of two $q\bar q$ dipoles starting at different times with the same color operator, which is denoted by straight black lines in \cref{fig:dijet} $(b)$. 
The resulting states are then projected onto each other after the evolution by identifying the quark and antiquark lines, which corresponds to inserting an appropriate projector between the two evolved states.
Note that this identification is also extended to momentum space. Thus, the correlators in \cref{eq:C4:C2:1} can also be written as
\begin{align}\label{eq:C4_C2_matrix_elements}
\begin{split}
    \mathcal{C}_4 & = \frac{1}{N_c}\big\langle\,\Tr \langle \bar\k_1 \bar\k_2 | U^\dagger_{q\bar q}(L,\bar x^+) \, \mathtt{M} \,U_{q\bar q}(L,x^+) | \k_1 \k_2 \rangle\,\big\rangle \;,\\
    \mathcal{C}_2 &  = \frac{1}{N_c}\big\langle\,\Tr \langle \p_1\p_2 | U_{q\bar q}(L,x^+)| \k_1 \k_2 \rangle\,\big\rangle \;,
\end{split}
\end{align}
where we defined the evolution operator 
\begin{align}\label{eq:Uqqbar}
    U_{q\bar q}(x_2^+,x_1^+) = G(x_2^+,x_1^+;p_1^+)\bar G(x_2^+,x_1^+;p_2^+) \, ,
\end{align}
for the $q\bar q$ system at fixed particle number and the propagators act on the momentum states to give 
\begin{multline}
   \hspace{0 cm} \langle \p_1,\p_2 |G(x_2^+,x_1^+;p_1^+)\bar G(x_2^+,x_1^+;p_2^+)|\k_1,\k_2\rangle = \\G(x_2^+,\p_1; x_1^+, \k_1; p_1^+)\bar G(x_2^+,\p_2, x_1^+, \k_2; p_2^+)\;.
\end{multline}

The relations in \cref{eq:C4_C2_matrix_elements} were conveniently written in such a way that the mapping to a quantum circuit is immediate, see \cref{fig:dijet}
 $(c)$. The contraction of the propagating dipoles corresponds to a projective measurement:
\begin{align}
  \mathtt{M} =  |\p_1 \p_2 \rangle \langle \p_1 \p_2 |   \, ,
\end{align}
which has an immediate quantum circuit formulation. Thus the full cross-section can be computed by preparing an initial state for the quark-antiquark pair (color singlet) and then unitarily evolving them in color and position spaces, before applying the projection. The evolution operators entering at this step have information about $\mathcal{A}$, and the quantum simulation has to be run for several samples of the field. The cross-section is obtained by making projective measurements on the quark and antiquark registers.

In the following section we shall present the realization of the above process following the form in \cref{eq:C4_C2_matrix_elements}. To this end, and aiming to have an analytical benchmark, we consider the case where the classical field $\mathcal{A}$ has Gaussian statistics
\begin{equation}\label{eq:A_correlator} 
    \langle \mathcal{A}_a^-(x^+,x^-,\boldsymbol{x})\mathcal{A}_b^-(y^+,y^-,\boldsymbol{y})\rangle = \delta_{ab}n(x^+,\x)\delta(x^+-y^+)\gamma(\x-\y)\, ,
\end{equation}
which are local in colour and light-cone time. The form of field correlations is then set by $\gamma(\x)$, which is directly related to the in-medium elastic scattering rate, and by the medium density $n(x^+,\x)$, which for a 
fixed length and homogeneous medium is given by $n(x^+,\x) = n(x^+)\,\Theta(x^+<L)$, and $L$ is usually of the order of a few fm. 
Given the symmetries of field correlations in \cref{eq:A_correlator}, let us now further simplify \cref{eq:gamma_qqbar_fullxsec}. Starting with the light-cone time locality, one can write $\cC_4$ as a momentum convolution of two objects evaluated at two distinct time domains
\begin{align}\label{eq:time_separation_C4}
\begin{split}
	\cC_4 = & \frac{1}{N_c^2}\int_{\q_1,\q_2} \big\langle\,\Tr \langle \bar\k_1 \bar\k_2 | U^\dagger_{q\bar q}(L,\bar x^+) \, \mathtt{M} \,U_{q\bar q}(L,\bar x^+) | \q_1 \q_2 \rangle\,\big\rangle \\
    & \times \big\langle\,\Tr \langle \q_1 \q_2 | U_{q\bar q}(\bar x^+,x^+) | \k_1 \k_2 \rangle\,\big\rangle \, ,
\end{split}
\end{align}
Further, the medium's translation invariance, i.e. the dependence on $\x-\y$ of the collision kernel $\gamma$ in \cref{eq:A_correlator}, implies that the total transverse momentum must be the same between amplitude and complex conjugate at all times~\cite{Isaksen:2023nlr}. Hence, one can simplify the total momentum dependence as
\begin{align}\label{eq:trans_inv_obj}
\begin{split}
   & \cC_4 \propto \delta^{(2)}(\BK - \bar\BK)e^{-i\frac{\BK^2}{2p_0^+}(\bar x^+ - x^+)}\hat{\cC}_4(\BK-\P, \,\cdots)\,, \\
   & \cC_2 \propto \delta^{(2)}(\BK-\P)e^{-i\frac{\BK^2}{2p_0^+}(L - x^+)}\hat{\cC}_2(\,\cdots)\,,
\end{split}
\end{align}
where the $\,\cdots$ stand for the remaining dependence on the relevant relative momenta for the placeholders $\hat \cC_4$ and $\hat \cC_2$. 
Furthermore, the dependence on the initial production amplitude can be factorized from the rest of the calculation by integrating over total momentum information $\diff^3\Omega_0$. Using these simplifications and re-writing the in-medium cross-section as 
\begin{align}\label{eq:Fmed_p_1}
	(2\pi)\frac{\diff\sigma}{\diff z\,\diff^2\p}  = (2\pi)\frac{\diff\sigma^{\rm vac}}{\diff z\,\diff^2\p} (1 + F_{\rm med}(\p^2,z))\;,
\end{align}
one arrives at the following expressions for the vacuum cross-section $\diff\sigma^{\rm vac}$ and for $F_{\rm med}$:
\begin{align}\label{eq:Fmed}
\begin{split}
     (2\pi)\frac{\diff\sigma^{\rm vac }}{\diff z\,\diff^2\p} = &N_c\frac{\alpha_e}{\pi}\frac{P_{q\gamma}(z)}{\p^2}\left(\sum_\lambda\int_{\Omega_0}\cM_0^\lambda(\BK,p_0^+)\cM_0^{\dagger,\lambda}(\BK,p_0^+)\right) \, ,\\
    F_{\rm med}(\p^2,z)  =\, & \frac{\p^2}{2\omega^2}{\rm Re}\int_0^L \diff x^+\int_{x^+}^L \diff\bar x^+\int_{\k_1,\k_2,\bar\k_1,\bar\k_2}(\bkappa \cdot \bar{\bkappa})e^{i\BK^2/2p_0^+(\bar x^+-x^+)}\left.\cC_4\right|_{\P=0}\\
	& -\frac{1}{\omega}{\rm Re}\,i\int_0^L \diff x^+\int_{\k_1,\k_2} (\bkappa\cdot\p)\left.\cC_2\right|_{\P=0} \;.
    \end{split}
\end{align}
Here, due to the form of \cref{eq:trans_inv_obj} and to the integration over total momentum, the subscript $\P=\p_1+\p_2=0$ constrains the contributions from the correlators $\cC_2$, $\cC_4$ to be diagonal in external momentum states, i.e., $\p_1=-\p_2=\p$. 
The object $F_{\rm med}$ quantifies the medium modification to the vacuum cross-section and its representation in terms of $\cC_2$ and $\cC_4$ is suitable for quantum simulation, where the matrix elements are evaluated at the amplitude level.

On the other hand, in terms of the explicit form of the in-medium propagator in \cref{eq:def_in_medium_propagator}, and following time locality and transverse homogeneity of \cref{eq:A_correlator}, the correlators $\cC_2$ and $\cC_4$ can be written as
\begin{align}\label{eq:Q_K_matrix_elements}
\begin{split}
    & \int_{\BK,\bar\BK}e^{i\BK^2/2p_0^+(\bar x^+-x^+)}\left.\cC_4\right|_{\P=0} = \int_\q\cQ(\p,\q,\bar\bkappa;L,\bar x^+)\cK(\q,\bkappa;\bar x^+,x^+) \,,\\
    & \int_\BK\left.\cC_2\right|_{\P=0} = \cK(\p,\bkappa;L,x^+) \, ,
\end{split}
\end{align}
where the objects $\cQ$ and $\cK$ read
\begin{align}\label{eq:Q_K_objects}
\begin{split}
   \cQ(\p,\q,\bar\bkappa;L,\bar x^+) =   & \int_{\u_1,\u_2,\bar \u_1, \bar\u_2} e^{i\u_1\cdot\q}e^{-i\bar\u_1\cdot \bar \bkappa}e^{-i(\u_2-\bar\u_2)\cdot\p}\\
    & \times \int_{\u_1}^{\u_2}\cD\u\int_{\bar \u_1}^{\bar \u_2}\cD\bar \u\, e^{\frac{i\omega}{2}\int_{\bar x^+}^{L}ds^+\, (\dot{\u}^2 - \dot{\bar \u}^2)}\frac{1}{N_c}\left\langle\Tr\left(U_1U_2^{\dagger}U_{\bar 2}U_{\bar 1}^{\dagger}\right)\right\rangle\;,\\
     \cK(\q,\bkappa;\bar x^+,x^+) = &\int_{\u_1,\u_2}e^{i\u_1\cdot\bkappa}e^{-i\u_2\cdot\q} \int_{\u_1}^{\u_2}\cD\u\, e^{\frac{i\omega}{2}\int_{x^+}^{\bar x^+}ds^+\,\dot{\u}^2}\frac{1}{N_c} \left\langle\Tr\left(U_1U_2^{\dagger}\right)\right\rangle
    \;,
     \end{split}
\end{align}
with $U_i\equiv U(\boldsymbol{r}_i)$, $\u = \r_1 - \r_2$, $\bar\u = \bar\r_1-\bar\r_2$, and $z(\r_1-\bar \r_1) + (1-z)(\r_2 - \bar\r_2) = 0$.
The kernel $\cK$ admits an analytical solution, while $\cQ$ generally requires numerical evaluation. A simple analytical expression of $\cQ$ can be obtained, however, if the so-called ``non-factorizable'' contributions are neglected, see e.g.~\cite{Blaizot:2012fh,Apolinario:2014csa,Dominguez:2019ges,Isaksen:2023nlr}. From numerical estimates, one expects this approximation to work reasonably well for small $\p^2/(p_0^+)^2$, i.e., soft and/or collinear $q\bar q$ pairs~\cite{Isaksen:2023nlr}. In this case, the
quadrupole operator factorizes as 
\begin{align}
    \cQ(\p,\q,\bar\bkappa;L,\bar x^+) \approx \delta^2(\q-\bar\bkappa)\int_\v e^{i(\p-\bar\bkappa)\cdot\v}\cP(\v;L,\bar x^+) \, ,
\end{align}
such that, in coordinate space, $F_{\rm med}$ reads
\begin{align}
\begin{split}
    \left(F_{\rm med}\right)^{\rm fctr.}(\p^2,z)
   =  &  \frac{\p^2}{2\omega^2}{\rm Re}\int_0^L \diff x^+\int_{x^+}^L \diff\bar x^+\int_{\v}e^{-i\p\cdot\v}\,\cP(\v;L,\bar x^+)(\partial_{\u_2} \cdot \partial_{\u_1})\\
    &\quad \times \cK(\u_2,\u_1;\bar x^+,x^+)\vert_{\u_1=0\,,\u_2=\v} \\
    &+\frac{1}{\omega}{\rm Re}\,\int_0^L \diff x^+\int_{\v} e^{-i\p\cdot\v}(\p\cdot\partial_{\u_1}) \cK(\v,\u_1; L,x^+)\vert_{\u_1=0}\,, 
\end{split} 
\end{align}
where the broadening and radiative kernels are given by, respectively,
\begin{align}
    & \cP(\v;L,\bar x^+) = \exp\left\{-\frac{C_F}{2}\int_{\bar x^+}^Lds^+\, n(s^+)\left(\sigma((1-z)\v) + \sigma(z\v)\right)\right\}\,,\nn
    & \cK(\u_2,\u_1;\bar x^+,x^+) = \int_{\u_1}^{\u_2}\cD\u\, e^{\frac{i\omega}{2}\int_{x^+}^{\bar x^+}ds^+\,\dot{\u}^2}\exp\left\{-\frac{C_F}{2}\int_{x^+}^{\bar x^+}ds^+ \,n(s^+)\sigma(\u)\right\}\,,
\end{align}
with the dipole potential $\sigma$ defined through the scattering rate in \cref{eq:A_correlator} by $\sigma(\r) = 2g^2(\gamma(0)-\gamma(\r))$. In the multiple soft scattering/harmonic oscillator regime, one approximates this potential by
\begin{align}
    C_F \int_{x^+}^{\bar x^+} ds \, n(s) \sigma(\boldsymbol{r}) 
    = (\bar x^+-x^+)\frac{\hat q \,\r^2}{2} + \mathcal{O}(\r^2\log\r^2) \, ,
\end{align}
where we further assume a static medium $n(x^+) = n_0$ and $\hat q = \hat q_F$, i.e., we work with the fundamental jet quenching parameter. Under these approximations (HO, fctr.), one can write~\cite{Isaksen:2023nlr}
\begin{align}\label{eq:Fmed_ana}
\begin{split}
     \left(F_{\rm med}\right)^{\rm HO,\, fctr.}(\p^2,z) 
     &=  -\frac{\p^2}{2\omega}\,2{\rm Re}\,i\int_0^L \diff x^+ \left(\frac{c_2}{c_3}\right)e^{\frac{i\p^2}{2\omega c_3}}- 2{\rm Re}\left(1-e^{-\frac{i\p^2\tan\Omega L}{2\omega\Omega}}\right)\;,
\end{split} 
\end{align}
where
\begin{align}\label{eq:ci_def}
     &c_{1}(x^+) = \frac{\Omega}{2i\sin{\Omega x^+}}\,, 
     \qquad  c_{2}(x^+) = \frac{\Omega}{\tan{\Omega x^+}}\,,\nn
     &c_{3}(x^+,\bar x^+) = x^+ \left(-i\frac{\hat q (z^2 + (1-z)^2)}{2\omega}\right)-c_{2}(\bar x^+)\,,
\end{align}
and their arguments in \cref{eq:Fmed_ana} are $c_2(x^+)$ and $c_3(L- x^+,\ x^+)$. 
The imaginary frequency reads $\Omega = \frac{1-i}{\sqrt{2}}\sqrt{\frac{\hat q}{2\omega}}$, 
$\omega = z(1-z)p_0^+$ and $p_0^+$ is the light-cone energy of the initial photon. Note that in the limit of $\hat q\rightarrow 0$ one has $F_{\rm med} \rightarrow 0$, resulting in $\diff\sigma = \diff\sigma^{\rm vac}$. 
For completeness, we also provide the analytic expressions for the momentum-integrated $\cC_2$ and $\cC_4$ objects below:
\begin{align}\label{eq:k_int_C2_C4}
\begin{split}
    \int_{\k_1,\k_2}(\bkappa\cdot\p)\left.\cC_2\right|_{\P = 0} 
     \stackrel{\rm HO,\, fctr.}{=} &
     \frac{\p^2}{\cos^2\Omega\Delta L}e^{-\frac{i\p^2\tan\Omega \Delta L}{2\omega\Omega} }\,,\\
    \int_{\k_1,\k_2, \bar \k_1, \bar \k_2} (\bkappa\cdot\bar\bkappa) e^{i\BK^2/2p_0^+(\bar x^+-x^+)}\left.\cC_4\right|_{\P=0} 
    \stackrel{\rm HO,\, fctr.}{=}&
    \,  2i\omega e^{\frac{i\p^2}{2\omega c_3}}\left(\frac{2ic_1}{c_3}\right)^2\\
    &\times \left(c_2 + c_3 + \frac{i\p^2}{2\omega}\frac{c_2}{c_3}\right)\,,
\end{split}
\end{align}
where the arguments of the $c_i$ functions are $c_1(\Delta x^+)$,  $c_2(\Delta x^+)$ and $c_3(\overline{\Delta L}, \Delta x^+)$, with $\Delta x^+ = \bar x^+ - x^+$, $\Delta L = L-x^+$ and $\overline{\Delta L} = L-\bar x^+$.

\subsection{Color (de)coherence of QCD antennas in nuclear matter}

\begin{figure}[t!]
    \centering
\begin{overpic}[width=0.7\textwidth]{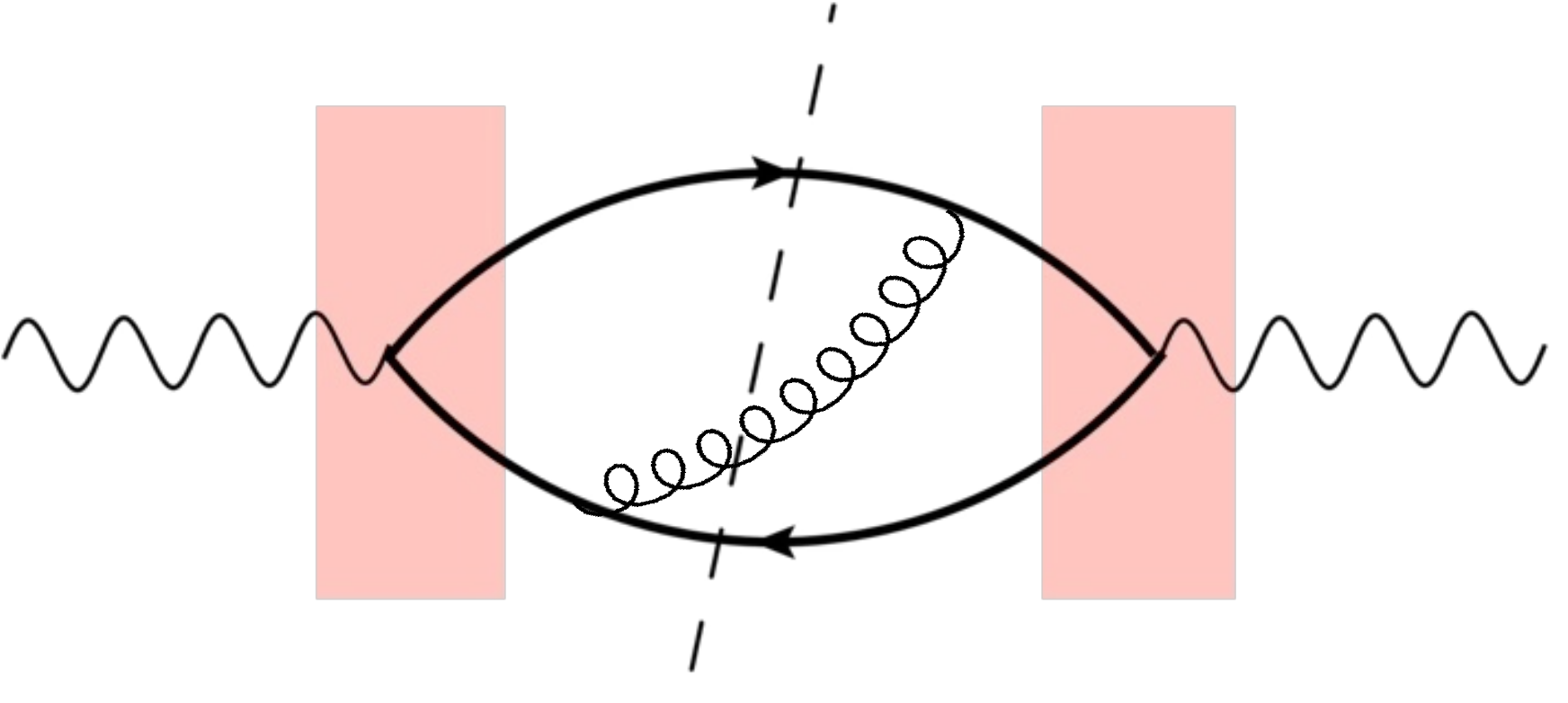}
    \put(10,26){$\boldsymbol{K}$}
    \put(23,27){$\k_1$}
    \put(23,13){$\k_2$}
    \put(40,36){$\p_1$}
    \put(58,34){$j$}
    \put(63,32){$i$}
    \put(33,8){$k$}
    \put(37,7){$m$}
    \put(60,6){$\p_2$}
    \put(72,27){$\bar\k_1$}
    \put(72,13){$\bar\k_2$}
    \put(85,26){$\boldsymbol{\bar K}$}
\end{overpic}
\caption{
\textbf{Leading order diagram from three particle production in the presence of a nuclear target.} 
The dashed line separates the amplitude from the conjugate amplitude. 
Particle momenta and color indices are labeled, and the background field is
indicated by the pink box.
}
\label{fig:Feynman_qqbarg}
\end{figure}
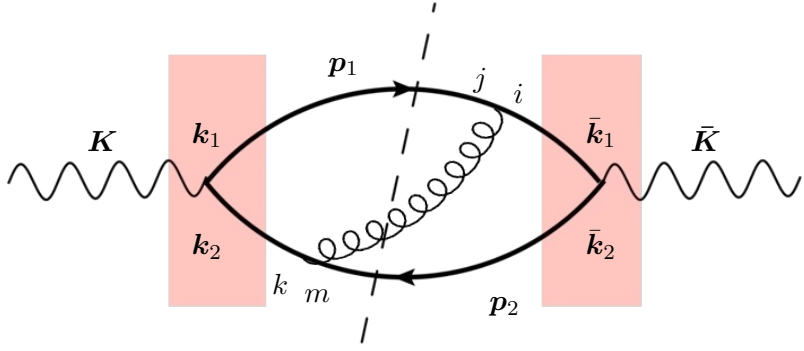

We now consider the diagram where there is a final state soft gluon being emitted off a high energy $q\bar q$ dipole produced from photon, as shown in \cref{fig:Feynman_qqbarg}. This set up has been considered in the past to study how color coherence is affected by the presence of the background matter~\cite{Mehtar-Tani:2010ebp,Mehtar-Tani:2011hma,Mehtar-Tani:2011vlz,Mehtar-Tani:2011lic,Casalderrey-Solana:2012evi,Mehtar-Tani:2017ypq,Barata:2021byj,Vaidya:2026yfa,Abreu:2024wka,Kuzmin:2025fyu,Andres:2025prc}, which can lead to the total loss of color correlation between the fermions -- our setup is very similar to~\cite{Abreu:2024wka} where a more complete discussion about the different coherence effects can be found. The gluon energy satisfies $k^+\ll p_1^+, p_2^+$, such that it is effectively emitted in vacuum. This picture is justified by the gluon's formation time being parametrically much larger than the medium length. Although the gluon's propagation is not modified by the medium, this process is relevant since the gluon acts as a measure of the color (de)coherence of the QCD antenna.

The calculation proceeds as described in \cref{subsec:dijet}, where we used the time-locality and translation invariance of field correlations in \cref{eq:A_correlator}. Combining both direct and interference terms on the same expression, while organizing it into contributions where the $q\bar q$ antenna splits in or out of the medium, one can write the cross-section in a convenient form:
\begin{subequations}
\begin{align}
\begin{split}
  \frac{(2\pi)\diff\sigma^{\rm in-in}}{\sigma_0 \diff z\diff  z_g \diff\theta_{qg}\diff^2\p}
   =  &
 \left[\frac{\alpha_s}{\pi}\frac{2C_F}{z_g\theta_{qg}} \frac{\alpha_e}{\pi}\frac{N_c P_{q\gamma}(z)}{ \p^2}\right]\frac{\p^2}{2\omega^2}{\rm Re}\int_0^L \diff x^+ \int_{x^+}^L \diff\bar x^+ \int_{\k_1,\k_2, \bar\k_1, \bar\k_2}\\
    & \times  (\bkappa\cdot\bar\bkappa)
   \left[\Theta(\theta_{qg} < \theta_{q\bar q})e^{i\frac{\BK^2}{2p_0^+}(\bar x^+-x^+)}\cC_4 
   \right.\\
   &\left.
   -\Theta(\theta_{qg} > \theta_{q\bar q})(\cC_{4,I}
    -e^{i\frac{\BK^2}{2p_0^+}(\bar x^+-x^+)}\cC_{4}) \right]_{\P=0}
    + {\rm sym.}
    \;,
    \end{split}\\
\begin{split}
    \frac{(2\pi)\diff\sigma^{\rm in-out}}{\sigma_0\,\diff z\,\diff z_g\, \diff\theta_{qg}\diff^2\p}
   = & \left[\frac{\alpha_s}{\pi}\frac{2C_F}{z_g\theta_{qg}} \frac{\alpha_e}{\pi}\frac{N_c P_{q\gamma}(z)}{ \p^2}\right]\Theta(\theta_{qg} < \theta_{q\bar q}){\rm Re}\,\left(\frac{-i}{\omega}\right)\int_0^L \diff x^+
    \\
    &\times \int_{\k_1,\k_2}\,(\bkappa\cdot\p)\left.\cC_2\right|_{\P=0} + {\rm sym.}\;,
       \end{split}\\
    \frac{(2\pi)^2\diff\sigma^{\rm out-out}}{\sigma_0\,\diff z\,d z_g\, \diff\theta_{qg}\diff^2\p}=  &\left[\frac{\alpha_s}{\pi}\frac{2C_F}{z_g\theta_{qg}} \frac{\alpha_e}{\pi}\frac{N_c P_{q\gamma}(z)}{ \p^2}\right]\Theta(\theta_{qg} < \theta_{q\bar q}) + {\rm sym.}\;,
\end{align}
\end{subequations}
where we have defined $z_g = k^+/p_1^+$ the energy fraction of the gluon with respect to the quark, $\theta_{qg}$ is the angle between the quark and the gluon, $\omega = z(1-z)p_0^+$, with $p_0^+$ the light-cone energy of the initial photon, while $z = p_1^+/p_0^+$ and $\p$ are, respectively, the energy fraction and the transverse momentum of the $q\bar q$ pair as defined in Section~\ref{subsec:dijet}. The angle $\theta_{q\bar q}$ is defined through 
$\p^2 = (z(1-z)E_0\,\theta_{q\bar q})^2$, where $E_0$ is the energy of the initial photon. Note that the subscript $\P=\p_1+\p_2=0$ constrains the contributions from the correlators $\cC_2$, $\cC_4$ and $\cC_{4,I}$ to be diagonal in external momentum states, i.e., $\p_1=-\p_2=\p$. 
The notation ``sym.'' denotes the corresponding contribution obtained by exchanging the roles of $q$ and $\bar{q}$.
Further, we have integrated over the azimuthal angle of the gluon with respect to the quark (to the anti-quark for the ``sym.'' contribution), such that we get a term with the usual angular ordering of soft gluon emissions and a pure medium-induced anti-angular ordered term manifested in $\Theta(\theta_{qg} > \theta_{q\bar q})$.
The angular-ordered contribution to the cross-section boils down to soft gluon emission factorized from the medium-modified $\gamma \to q\bar q$ cross-section, which involves the matrix elements $\mathcal{C}_4$ and $\mathcal{C}_2$ defined previously in \cref{eq:C4_C2_matrix_elements}. The anti-angular ordered contribution arises from the additional $t^a$ matrix from gluon emission, which alters the color structure of the interference term. This is encoded in a new correlator that reads
\begin{align}\label{eq:C4I_matrix_elements}
\begin{split}
    & \mathcal{C}_{4,I} = \frac{1}{C_FN_c}\big\langle\,\Tr \langle \bar\k_1 \bar\k_2 | U^\dagger_{q\bar q}(L,\bar x^+) \, (t^a\otimes \mathtt{M}\otimes t^a) U_{q\bar q}(L,x^+) | \k_1 \k_2 \rangle\,\big\rangle 
    \;.
\end{split}
\end{align}
Here we use a form suitable for implementation in a quantum circuit, where the tensor product denotes the action of the color matrices on the color registers for the quark and antiquark. In terms of the $\cK$ emission kernel defined in \cref{eq:Q_K_objects}, this correlator reads\footnote{
Note that in the large-$N_c$ limit, the correlator $\cC_{4,I}$ is free of quadrupoles, unlike $\cC_4$ (see  \cref{eq:Q_K_objects}).
}
\begin{align}
    \int_{\k_1,\k_2,\bar\k_1,\bar\k_2}(\bkappa\cdot\bar\bkappa)\left.\cC_{4,I}\right|_{\P = 0} = \int_{\bkappa,\bar\bkappa}(\bkappa\cdot\bar\bkappa) \cK(\p, \bkappa; L, x^+)\cK^{\ast}(\p, \bar\bkappa; L, \bar x^+) 
\end{align}
Defining $F_{\rm med,\,g}$ as the ratio between the pure medium-induced contribution to the full cross-section and the angular ordered vacuum cross-section
\begin{align}
    & F_{\rm med,\,g}(\p^2, z) = \left(\diff\sigma^{\rm in-in} + \diff\sigma^{\rm in-out}\right)\Big / \left(\frac{\alpha_s}{\pi}\frac{2C_F}{z_g\theta_{qg}} \frac{\alpha_e}{\pi}\frac{N_c P_{q\gamma}(z)}{ \p^2}\right) \, ,
\end{align}
then one can write
\begin{subequations}
\begin{align}
     \Theta(\theta_{qg} < \theta_{q\bar q})F_{\rm med,\,g}(\p^2, z) = & \Theta(\theta_{qg} < \theta_{q\bar q})F_{\rm med}(\p^2,z)\;,\\
\label{eq:F_med_anti_ang}
\begin{split}
    \Theta(\theta_{qg} > \theta_{q\bar q})F_{\rm med,\,g}(\p^2, z)   =& -\Theta(\theta_{qg} > \theta_{q\bar q})\frac{\p^2}{2\omega^2}{\rm Re}\,\int_0^L \diff x^+ \int_{x^+}^L \diff\bar x^+ \int_{\k_1,\k_2, \bar\k_1, \bar\k_2} \\
   &\times (\bkappa\cdot\bar\bkappa)(\cC_{4,I}-e^{i\frac{\BK^2}{2p_0^+}(\bar x^+-x^+)}\cC_{4})\big|_{\P=0}\;.
\end{split}
\end{align}
\end{subequations}
Note that only the anti-angular ordered part is of interest, since the angular ordered contribution  exactly matches the object $F_{\rm med}(\p^2,z)$ analyzed in the previous section. Following the same assumptions as used for the dipole formation, namely keeping only the factorized contribution to $\cC_4$ and using the multiple soft scattering/harmonic oscillator approximation, we can write it analytically as
\begin{align}\label{eq:F_med_anti_ang_ana}
\begin{split}
     & \Theta(\theta_{qg} > \theta_{q\bar q}) \left(F_{\rm med,\,g}\right)^{\rm HO,\,fctr.}(\p^2, z)= \\
    & 
     -\Theta(\theta_{qg} > \theta_{q\bar q})\left[\frac{\p^2}{2\omega}2{\rm Re}\,i\int_0^L \diff x^+ \left(\frac{c_2}{c_3}\right)
     e^{\frac{i\p^2}{2\omega c_3}}
     + \left|1 - e^{-\frac{i\p^2 \tan\Omega L}{2\omega \Omega}}\right|^2\right]\;,
\end{split}
\end{align}
with the functions $c_i$ defined in \cref{eq:ci_def} and their arguments are $c_2(x^+)$ and $c_3(L-x^+,x^+)$.
As presented in \cref{eq:k_int_C2_C4} for $\cC_2$ and $\cC_4$, the analytic result for the momentum integrated $\cC_{4,I}$ reads
\begin{align}\label{eq:k_int_C4_C4I}
\begin{split}
    \int_{\k_1,\k_2,\bar\k_1,\bar\k_2} (\bkappa\cdot\bar\bkappa)\left.\cC_{4,I}\right|_{\P = 0}
    \stackrel{\rm HO,\, fctr.}{=}
    & 16\p^2
    \left(\frac{c_1(\Delta L)}{c_2(\Delta L)}\right)^2
    \left(\frac{c_1(\overline{\Delta L})}{ c_2(\overline{\Delta L})}\right)^{2,\ast}\\
    &
    \times \exp\left\{-i\frac{\p^2}{2\omega}\left(\frac{1}{c_2(\Delta L)} - \frac{1}{ c_2^{\ast}(\overline{\Delta L})}\right)\right\}\;.
\end{split}
\end{align}

We have thus expressed the medium modifications $F_{\rm med}$ and $F_{\rm med,\,g}$ in terms of the matrix elements (see \cref{eq:Fmed,eq:F_med_anti_ang}) --- $ \mathcal{C}_2$, $ \mathcal{C}_4$, and $ \mathcal{C}_{4,I}$ --- that can be evaluated via quantum simulation. The implementation of this procedure will be discussed in the following section.
For comparison, we have also derived the corresponding analytical expression under additional assumptions, given in \cref{eq:Fmed_ana,eq:F_med_anti_ang_ana}. 
Comparing the quantum simulation results with these analytical expressions will therefore serve two purposes: it provides a check of the quantum implementation in the relevant limiting cases and tests the validity of the additional assumptions.

\section{Realization as a quantum circuit  }\label{sec:method}

In this section we detail the computation of the medium modification factors $F_{\rm med}$ and $F_{\rm med, g}$ via quantum simulation methods.
Specifically, we map the contributing matrix elements, $ \mathcal{C}_2$, $ \mathcal{C}_4$ and $ \mathcal{C}_{4,I}$, to unitary evolutions of quantum states on a quantum circuit.
The unitary evolution is realized by first constructing the corresponding light-front Hamiltonian in a chosen basis representation, and then evolving the quantum state on the circuit.  

We follow the scalable quantum simulation framework introduced in~\cite{Qian:2024gph}, which employs a direct encoding scheme implemented on quantum devices (see also the theoretical developments in \cite{Li:2020uhl,Li:2021zaw,Li:2023jeh,Li:2025wzq} and other quantum simulation algorithms in \cite{Barata:2022wim,Barata:2023clv,Barata:2021yri,Wu:2024adk,Castro:2025ocx}). 
In that work, the framework was applied to quark and gluon jets, and quantum states evolve from given initial states and are measured at the final states. 
Here, we formulate the problem in the valence Fock sector of the quark–antiquark dipole, i.e., $\ket{q\bar{q}}$, and the full process is more involved, consisting of a forward time evolution, a projection, and a backward time evolution. 

\subsection{Light-front Hamiltonian}
The light-front QCD Hamiltonian for a quark-antiquark system in the presence of a gluonic background field $ \mathcal A$ consists of kinetic energy and interaction terms with the field,
\begin{align}
\begin{split}
    P^-(x^+) &= P^-_\mathrm{KE} + V_\mathcal{A}(x^+)\\
    &=\left(P^-_{\mathrm{KE},q} + V_{\mathcal{A},q}(x^+)\right)+\left(P^-_{\mathrm{KE}, \bar{q}} + V_{\mathcal{A},\bar{q}}(x^+)\right)\;.
\end{split}
\end{align}

We adopt a discrete momentum basis representation, following the setup of ~\cite{Qian:2024gph,Li:2021zaw,Li:2023jeh,Li:2025wzq,Barata:2023clv,Barata:2022wim}.
We consider a system contained within a finite box of volume $L^-(2L_\perp)^2$, where \(L^-\) is the extent in the longitudinal direction $x^-$ and \(2L_\perp\) is that in the transverse directions. 
The transverse coordinates span the range $-L_\perp \le r^i\le L_\perp$. 
Imposing periodic boundary conditions, the transverse coordinates are discretized as 
  \begin{align}
    r^i= n_i a_\perp \quad (i = x,y)  \;,
 \end{align}
where $n_i=-N,-N+1,\ldots,N-1 $ and the lattice spacing is $a_\perp=L_\perp/N_\perp$. 
  The corresponding momentum space is likewise discrete with periodic boundary conditions,
  \begin{align}
     & p^i = k^i d_p \;,
  \end{align}
where $k^i = -N_\perp, -N_\perp+1 \ldots, N_\perp-1$ and $d_p\equiv \pi/L_\perp$ is the resolution in momentum space. 

The mode expansion of the field operators in such a discrete momentum basis is
\begin{align}
    \begin{split}
    \Psi_{\mu} (x)=  &\sum_{\beta} \frac{1}{\sqrt{ p^+ L^- (2L_\perp)^2}}
    \left[b^\pd_{\ind} u_\mu(p,\lambda)
    e^{-ip\cdot x}
    +d^\dagger_{\ind} v_\mu(p,\lambda) e^{ip\cdot x}\right]\;, 
    \end{split}
\end{align}
where $p\cdot x=(1/2)\,p^+ x^- -\p \cdot \x$ is the 3-product for the spatial components of $p^\mu$ and $x^\mu$. Each single-particle state is specified by five quantum numbers, 
\begin{align}
  \ind=\{ k^+, k^x, k^y, \lambda,c \}\;,
\end{align}
where $\lambda$ denotes the light-front helicity and $c=1,2,\ldots, N_c$ is the color index of the fermion.
Since the Hamiltonian conserves $k^+$ and $\lambda$, only $ \{ k^x, k^y, c \}$ remain as the nontrivial quantum numbers throughout the evolution. 
The creation operators $b^{\dagger}_\ind$ and $d^{\dagger}_\ind$ create quarks and antiquarks with quantum numbers $\ind$ respectively. 
They obey the following commutation relations,
\begin{align}\label{eq:commutations}
    \begin{split}
        & \{b^\pd_\ind,b^{\dagger}_{\ind'}\}=\{d^\pd_{\ind},d_{\ind'}^{\dagger}\}
        =\delta^\pd_{\ind,\ind'} \;,  
    \end{split}
\end{align}
with all the other anti-commutators vanishing.
Accordingly, the single-particle basis states are defined as 
\begin{align}
    \begin{split}
       &\ket{q(\ind)}
    =b^\dagger_{\ind}\ket{0}\;,\\
           &\ket{\bar q(\ind)}
    =d^\dagger_{\ind}\ket{0}\;.
    \end{split}
\end{align}
The multiple-particle basis states are constructed as tensor products of the single-particle states, e.g., the basis state of a dipole is $ \ket{q (\ind)\bar q(\ind')}=\ket{q(\ind)}\otimes \ket{\bar q(\ind')}$. 
The dipole state in the $\ket{q\bar{q}}$ Fock sector can then be expanded as
\begin{align}
    \ket{\psi_{q\bar q}}=\sum_\beta \sum_{\beta'} c_{\beta \beta'} \ket{q (\ind)\bar q(\ind')}\;.
\end{align}
The second-quantized operators in this basis can be written explicitly as~\cite{Qian:2024gph} 
\begin{subequations}
\label{eq:secondquantized}
    \begin{align}
        & P^-_{KE,q}= \sum_{\ind} \frac{\p^2+m_q^2}{p^+} b_\ind^\dagger b^\pd_\ind\;,\\
        &  P^-_{KE,\bar{q}}=\sum_{\ind} \frac{\p^2+m_{ q}^2}{p^+}d_\ind^\dagger d^\pd_\ind\;,\\
        & V_{q\mathcal{A}} (x^+)
        =\frac{2g}{(2L_\perp)^2 }\sum_{\ind_1,\ind_2} 
        \delta_{k_2^+,k_1^+}
        \delta_{\lambda_1,\lambda_2}
         T_{c_2,c_1}^a
         \tilde{\mathcal{A}}^a_+(\p_2- \p_1, x^+)
         b^\dagger_{\ind_2} 
        b^\pd_{\ind_1} 
         \;,\\
        & V_{\bar q\mathcal{A}} (x^+)
        =-\frac{2g}{(2L_\perp)^2 }\sum_{\ind_1,\ind_2} 
        \delta_{k_2^+,k_1^+}
        \delta_{\lambda_1,\lambda_2}
         T_{c_1,c_2}^a
         \tilde{\mathcal{A}}^a_+(\p_2- \p_1, x^+)
         d^\dagger_{\ind_2} 
        d^\pd_{\ind_1} 
        \;.
    \end{align}
\end{subequations}
The background field $\mathcal A$ is formulated in the same transverse basis space as the fermion fields, and is obtained by solving the reduced Yang-Mills equation
\begin{align}\label{eq:poisson}
 (m_g^2-\nabla^2_\perp )  \mathcal{A}^-_a(\x,x^+)=\rho_a(\x,x^+)\;,
\end{align}
where $m_g$ is the effective gluon mass of the field introduced to regularize the infrared divergence.
The color charge density source $\rho_a$ satisfies the local correlation function
\begin{align}\label{eq:MV_color_charge}
    \langle \rho_a(x^+,\x)\rho_b({x'}^{+},\x') \rangle =
    g^2 n(x^+,\x)
    \delta_{ab}\,\delta^{(2)}(\x-\x')\,\delta(x^+-{x'}^+)\;.
\end{align}
In the discrete basis space, this correlation function takes the form~\cite{Lappi:2007ku, Li:2020uhl, Li:2021zaw}:
\begin{align}\label{eq:chgcor_dis}
    \braket{\rho_a(n_x,n_y,n_\tau)\rho_b({n'}_x,{n'}_y,n_\tau')}
    =g^2 
    n(x^+,\x)
    \delta_{ab}\frac{\delta_{n_x,{n'}_x}\delta_{n_y,{n'}_y}}{a_\perp^2}\frac{\delta_{n_\tau,n_\tau'}}{\tau}\;,
\end{align}
where \(\x = (n_x, n_y) a_\perp\) and \(x^+ = n_\tau \tau \), with the layer index running over \( n_\tau = 1, 2, \ldots, N_\eta \), and $N_\eta$ denoting the total number of layers of the color source along $x^+ \in [0,L]$.
At the amplitude level, the color charge density \(\rho\) is a stochastic variable drawn from a Gaussian with zero mean and variance \( g^2  n(x^+,\x) / (a_\perp^2 \tau) \) at each lattice site \( \{ n_x, n_y, n_\tau \} \).
This distribution allows for the numerical sampling of \(\rho\) at each site and for each event.
For a static and homogeneous medium, we take $n(x^+,\x)=n_0\,\Theta(x^+<L^+)$, and the corresponding background field is obtained directly by solving \cref{eq:poisson}.
The transverse profile function $\gamma(\r)$ appearing in the field correlator~\cref{eq:A_correlator} is then
\begin{align}\label{eq:gamma}
    \gamma(\r) = \int \frac{d^2\q}{(2\pi)^2} e^{i\q\cdot \r}\frac{g^2}{(\q^2+m_g^2)^2} = g^2\frac{\r K_1(m_g \r)}{4\pi m_g}\;.
\end{align}
The quenching parameter follows as~\cite{Barata:2022wim, Li:2023jeh} 
\begin{align}\label{eq:qhat}
  \hat q
 =
 \frac{C_F
 g^4 n_0 }{4\pi}
  \biggl[
    \log\left(1+\frac{1}{(m_g a_\perp/\pi)^2}\right)
    -\frac{1}{1+(m_g a_\perp/\pi)^2}
  \biggr] \;.
\end{align}
In the HO approximation, where the logarithmic and power corrections in the brackets are neglected, this reduces to 
\begin{align}\label{eq:qhat0}
  \hat q_0
\equiv
 \frac{C_F
 g^4 n_0}{4\pi}\;,
\end{align}
which matches the definition of $\hat q$ used in analytical result in Section~\ref{subsec:dijet}.

\subsection{Evaluation of matrix elements}

The time evolution of the quantum state is governed by
\begin{align}
  \label{eq:ShrodingerEq}
  i\frac{\partial}{\partial x^+}\ket{\psi;x^+}=\frac{1}{2}P^-(x^+)\ket{\psi;x^+}\;.
\end{align}
The evolved state can be written in terms of the evolution operator,
\begin{align}\label{eq:ShrodingerEqSol}
  \begin{split}
\ket{\psi;x^+}&=U_{q\bar q}(x^+,0)\ket{\psi;0}\,,
\qquad
U_{q\bar q}(x^+_f,x_i^+)\equiv\mathcal P \exp\left[-\frac{i}{2}\int_{x_i^+}^{x^+_f}\diff z^+P^-(z^+)\right]\;.
  \end{split}
\end{align}
Note that the definition of $U_{q\bar q}$ is equivalent to that in \cref{eq:Uqqbar}.\footnote{Note that the sign convention for the exponential in the path ordered Wilson line \cref{eq:def_wilson_line} in does not match the convention for the evolution operator in~\eqref{eq:ShrodingerEqSol}. This, however, does not affect the final result.} 

We implement the time evolution by first Trotterizing the evolution operator $U_{q\bar q}$ into $N_s$ successive time steps, 
\begin{align}\label{eq:Uqqbar_trotter}
  \begin{split}
    U_{q\bar q}(x^+,0)
    &=
    \prod^{N_s-1}_{k=0}
    U_{q\bar q}(x_{k+1}^+,x_{k}^+)\,,
    \qquad 
    x_k^+=k\delta x^+\,, \text{with } \delta x^+=x^+/N_s
    \;.
  \end{split}
\end{align}
We take $N_s$ sufficiently large, i.e., the timestep $\delta x^+$ sufficiently small, and apply a product formula\footnote{In our setup, when the time step is within one layer of the field, $\delta x^+\leq\tau$, the Hamiltonian is already constant over the step. Nevertheless, the error of the product formula scales as $\mathcal O((\delta x^+)^2)$, so in practice one must take $\delta x^+ $ much smaller than $\tau$ to achieve numerical convergence. }
\begin{align}
 \begin{split}\label{eq:trotter_1st}
   U_k\equiv &\lim_{\delta x^+\to 0}U_{q\bar q}(x_{k+1}^+,x_{k}^+)
  =\exp\left[-\frac{i}{2} P_{KE}^-\delta x^+\right]
  \exp\left[-\frac{i}{2} V_{\mathcal A}(x_k^+) \delta x^+\right]
  \;.
 \end{split}
\end{align}
The full evolution is thus given by the ordered product $\prod^{N_s-1}_{k=0} U_k$, which can be implemented directly as a sequence of unitaries on a quantum circuit.
The simulation results depend only on the following dimensionless combinations and are invariant under parameter transformations that keep these quantities fixed:
\begin{align}\label{eq:paras}
           \left\{\dfrac{L }{ a_\perp^2 p_0^+}, z,\, g^2 a_\perp \sqrt{n_0  L}, N_\eta, m_g a_\perp \right\}\;.
\end{align}

Using the implementation of the time-evolution operator, we can solve for the matrix elements $\mathcal{C}_2$, $\mathcal{C}_4$, and $\mathcal{C}_{4,I}$ in \cref{eq:C4_C2_matrix_elements} by combining the forward- and backward-evolved states with the appropriate projectors.
For convenience, we define a color-singlet state with definite momentum as
\begin{align}\label{eq:psi_s}
    \ket{\psi_s(\p_1,\p_2)} \equiv&  \frac{1}{\sqrt{N_c}}\sum_{c=1}^{N_c}\ket{q(\p_1, c) \bar q(\p_2, c)}\;.
\end{align}
Here $c$ labels the color index, with the same value $c$ appearing for both the quark and antiquark enforcing the color-singlet condition. 
Here and in what follows, we keep $k^+$ and $\lambda$ implicit for notational simplicity.
For given splitting times $x^+,\bar x^+$ and momenta, the forward- and backward-evolved states are
\begin{align}\label{eq:psi_RL}
    \ket{\psi_R} =U_{q\bar{q}}(L,x^+) \ket{\psi_s(\k_1,\k_2)}\;,
\qquad
    \ket{\psi_L} =  U_{q\bar{q}}(L,\bar x^+)  \ket{\psi_s(\bar \k_1,\bar \k_2)}\;.
\end{align}
The matrix elements then take the form
\begin{align}\label{eq:C4_C2_matrix_elements_RL}
    \begin{split}
        &\mathcal{C}_2   = \expconfig{  
            \braket{\psi_s(\p_1,\p_2) |\psi_R}
            }\,, \\
          &  \mathcal{C}_4  = \expconfig{  
            \braket{\psi_L |
            (I\otimes \mathtt{M}\otimes I )
            |\psi_R}
            }\,, \\
        &\mathcal{C}_{4,I}  = \frac{1}{C_F}
        \expconfig{  
            \braket{\psi_L |(t^a\otimes \mathtt{M}\otimes t^a)|\psi_R}
            }
        \;.
    \end{split}
\end{align}
Recall that the measurement function is defined as $\mathtt{M} = |\p_{1},\p_2\rangle\langle\p_{1},\p_2|$. In the expression for $\mathcal{C}_4$, we explicitly restore the color-space projectors, which is just the identity operators acting on the quark and antiquark, respectively. Written explicitly, the projectors are
\begin{align}\label{eq:projectors}
\begin{split}
    & I\otimes \mathtt{M}\otimes I 
    =\sum_{i,j}|q(\p_{1},i),\bar q(\p_2,j)\rangle
    \langle q(\p_{1},i),\bar q(\p_2,j)|\;,\\
   & t^a\otimes \mathtt{M}\otimes t^a 
    =\sum_{i,j,k,m}t^a_{km} t^a_{ij} |q(\p_{1},i),\bar q(\p_2,m)\rangle
    \langle q(\p_{1},j),\bar q(\p_2,k)|\;.
    \end{split}
\end{align}
Since the background field $\mathcal{A}$ is stochastic, the quantum expectation values must be averaged over an ensemble of background-field configurations, as indicated by $\expconfig{\cdots}$ in \cref{eq:C4_C2_matrix_elements_RL}, consistent with the notation introduced in \cref{eq:C4:C2:1}.

After obtaining all matrix elements, the final $F_\mathrm{med}$ can be evaluated according to \cref{eq:Fmed} for collinear dipole formation, and $F_{\rm med, g}$ according to \cref{eq:F_med_anti_ang} with the additional emitted gluon.
Note that the these medium modifications involve time integrations of the matrix elements. 
Since the time evolution is performed with a finite number of
discretized time steps, we evaluate the integrations using Gauss--Legendre quadrature after applying cubic interpolation on the matrix elements in time.

Since the simulation is performed on a square lattice with periodic boundary conditions, two issues must be addressed to control lattice
artifacts. First, momentum conservation, $\P=\p_1+\p_2$, must be implemented consistently with the periodic boundary conditions, following the prescription detailed in Appendix C of 
\cite{Li:2021zaw}. Second, the transverse lattice is not perfectly symmetric around $k^i=0$ $(i=x,y)$, since it spans from $-N_\perp$ to $N_\perp-1$. As a result, the symmetry of the modification factor around
$z=1/2$ may be violated, particularly for small lattices. While the relation $F_{\rm med}(z,\p)=F_{\rm med}(1-z,-\p)$ remains exact on the lattice, the symmetry in $z$ alone is not guaranteed.
To compensate this artifact, we symmetrize the result as
\[
F_{\rm med}(z,\p)=\frac{1}{2}\left[F_{\rm med}(z,\p)
+F_{\rm med}(1-z,\p)\right].
\]

\subsection{Towards the implementation in a quantum device}

To implement the quantum simulation on the circuit, we use direct encoding scheme for the Hamiltonian operator where each qubit is used to encode the occupation mode of the fermions or antifermions~\cite{Qian:2024gph}. 
In doing so, the basis index of each particle state $\beta$ is uniquely mapped to some qubit register index $j$.
Then, the finite-dimensional fermionic Fock space can be represented by a multi-qubit system where the total number of qubits $N_q=2N_F$ for the $N_F$ many fermions and equally many anti-fermions. Since our momentum space is discretized from $-N_\perp$ to $N_\perp-1$ for both transverse directions, the total number of qubits $N_q = 2N_F = 2 N_c(2N_\perp)^2=8N_c N_\perp^2$ for the main circuit of the evolution. Because $k^+$ and $\lambda$ are conserved by $P^-(x^+)$ on a particle-by-particle basis, they enter the qubit formulation as fixed parameters rather than dynamical degrees of freedom, and so do not contribute to the qubit count.
The fermionic creation and annihilation operators are defined as
\begin{subequations}
\label{eq:fops}
\begin{align}\label{eq:fermionic_encoding}
 &\begin{multlined}
    b^\dag_j\ket{1^{\occf_1}, \ldots, j^{\occf_j}, \ldots, N_F^{\occf_{N_F}}} = (-1)^{\sum_{j<i} \occf_j}
    \sqrt{1-j}\ket{1^{\occf_1}, \ldots, (1-j)^{\occf_j}, \ldots, N_F^{\occf_{N_F}}},
 \end{multlined}\\
 &\begin{multlined}
    b_j\ket{1^{\occf_1}, \ldots, j^{\occf_j}, \ldots, N_F^{\occf_{N_F}}} =(-1)^{\sum_{j<i} \occf_j}
    \sqrt{j}\ket{1^{\occf_1}, \ldots, (1-j)^{\occf_j}, \ldots, N_F^{\occf_{N_F}}}\;,
 \end{multlined}
\end{align}
\end{subequations}
where $\occf_j$ is the occupancy of the $j$\textsuperscript{th} fermionic mode. To account for the anti-commutativity of fermionic creation and annihilation operators in \cref{eq:commutations}, one can use the standard Jordan-Wigner (JW) fermion-to-qubit mappings~\cite{Jordan:1928wi},
\begin{align}
\label{eq:JW}
&b_j^\dagger \mapsto \bigg(\prod_{k=1}^{j-1}Z_{k}\bigg)\otimes \frac{X_j-iY_j}{2}\;,\quad b_j \mapsto \bigg(\prod_{k=1}^{j-1}Z_{k}\bigg)\otimes \frac{X_j+iY_j}{2}\;, 
\end{align}
where $X_j, Y_j, Z_j$ are the Pauli-X, Pauli-Y, and Pauli-Y matrix on qubit $j$. The same procedure is carried out for the anti-fermion operators over all the qubits. However, as seen in eq.~\eqref{eq:secondquantized}, the Hamiltonian is a sum of terms that act on the quark or antiquark separately, with the other species as a spectator. The full evolution operator therefore factorizes into quark and antiquark blocks, and the cross-anticommutator $\{b_\beta, d_{\beta'}\} = 0$ is never invoked, and we can adopt the simpler two-block encoding on separate qubit registers. Additionally, since our Hamiltonian $P^-(x^+)$ strictly conserves $\ket{q\bar{q}}$ Fock sector, we can use Jordan-Wigner encoding without requiring $\mathcal{O}(N_F)$ Pauli-Z strings, greatly simplifying the computational cost. Then, the full Hamiltonian contains only $\mathcal{O}(N_F^2)$ Pauli terms. This direct encoding allows us to efficiently represent the Hamiltonian operator and the quantum state in the $\ket{q\bar{q}}$ Fock sector.

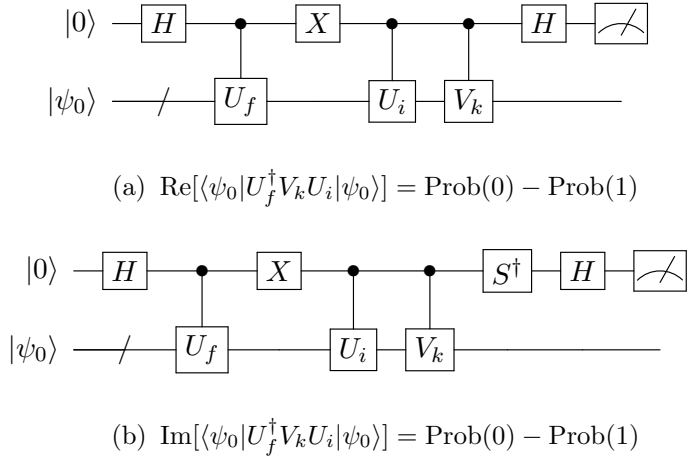
\begin{figure}[t]
\centering
\begin{subfigure}{1\textwidth}
\centering
\[
\Qcircuit @C=1.0em @R=1.2em {
\lstick{\ket{0}}   & \gate{H} & \ctrl{1} & \gate{X} & \ctrl{1} & \ctrl{1} & \gate{H} & \meter \\
\lstick{\ket{\psi_0}} & \qw/\qw      & \gate{U_f} & \qw     & \gate{U_i} & \gate{V_k} & \qw & \qw
}
\]
\subcaption{\,$\mathrm{Re}[\langle \psi_0 | U_f^\dagger V_k U_i | \psi_0 \rangle] = \mathrm{Prob}(0) - \mathrm{Prob}(1)$
\label{fig:hadamard-circuit-real}
}
\end{subfigure}
\vspace{1em}
\begin{subfigure}{1\textwidth}
\centering
\[
\Qcircuit @C=1.0em @R=1.2em {
\lstick{\ket{0}}   & \gate{H} & \ctrl{1} & \gate{X} & \ctrl{1} & \ctrl{1} & \gate{S^\dagger} & \gate{H} & \meter \\
\lstick{\ket{\psi_0}} & \qw/\qw     & \gate{U_f} & \qw     & \gate{U_i} & \gate{V_k} & \qw & \qw & \qw
}
\]
\subcaption{\,$\mathrm{Im}[\langle \psi_0 | U_f^\dagger V_k U_i | \psi_0 \rangle] = \mathrm{Prob}(0) - \mathrm{Prob}(1)$
\label{fig:hadamard-circuit-imag}
}
\end{subfigure}
\caption{
Quantum circuits to extract the real (top) and imaginary (bottom) parts of the matrix elements associated with a Pauli string $V_k$. 
Here $\ket{\psi_0} \equiv \ket{\psi_s(\p_1,\p_2)}$ for brevity.
\label{fig:hadamard-circuit}
}
\end{figure}

We now describe one possible approach to evaluate the matrix elements $\mathcal{C}_2$, $\mathcal{C}_4$, and $\mathcal{C}_{4,I}$ directly on the quantum circuit.
As shown in \cref{eq:C4_C2_matrix_elements_RL}, each of them can be evaluated by the same generic form
$\braket{\psi_f| \mathrm{P} |\psi_i}$, where the initial states $\ket{\psi_i}$ and final states $\ket{\psi_f}$ are generated by unitary evolution $\ket{\psi_f} = U_f \ket{\psi_s(\p_1,\p_2)}$ and $\ket{\psi_i} = U_i \ket{\psi_s(\p_1,\p_2)}$.\footnote{For all three matrix elements, $U_i=U_{q\bar{q}}(L,t)$. For $\mathcal{C}_2$, $U_f=I$, while for $\mathcal{C}_4$ and $\mathcal{C}_{4,I}$, $U_f=U_{q\bar{q}}(L,t')$.
The projector $\mathrm{P}$ takes $I$, $I\otimes \mathtt{M}\otimes I$, and $t^a\otimes \mathtt{M}\otimes t^a/C_F$ for $\mathcal{C}_2$, $\mathcal{C}_4$, and $\mathcal{C}_{4,I}$ respectively.}
For the initial state, we encode $\ket{\psi_s(\p_1,\p_2)}$ defined in \cref{eq:psi_s} straightforwardly using our direct encoding scheme by preparing the superposition state with $X$ and Hadamard gates acting on the appropriate qubits. For the time evolution, since $U_i$ and $U_f$ are unitary operators, they can be implemented as the standard quantum circuits for Trotterized time evolution. Lastly, the real part and imaginary part of the overlap can be obtained using the modified Hadamard test illustrated in \cref{fig:hadamard-circuit}, which requires an additional ancillary qubit~\cite{Cleve:1997dh}. 

For all three matrix elements in our calculations, the $\mathrm{P}$ operators are sparse Hermitians with only a constant number of nonzero elements, up to 4 terms in the case of $\mathcal{C}_{4,I}$. As P is not necessarily unitary (such as the case of $\mathcal{C}_{4,I}$), we need to decompose $\mathrm{P}$ into a linear combination of unitary Pauli strings $V_k$, i.e., $\mathrm{P} = \sum_k c_k V_k$, such that the total overlap $\langle \psi_f | \mathrm{P} | \psi_i \rangle$ is re-evaluated classically by summing the weighted expectation values $\sum_k c_k \langle V_k \rangle$. Following \cref{fig:hadamard-circuit-real}, taking the evaluation of the real part of the overlap for example, the final state of the composite system before the measurement is
\begin{align}
   \ket{\psi} = \frac{1}{2}\ket{0}\otimes \big(V_k \ket{\psi_i} + \ket{\psi_f}\big) + \frac{1}{2}\ket{1}\otimes \big(-V_k \ket{\psi_i} + \ket{\psi_f}\big)\,,
\end{align}
where respective measurements on the ancilla give
\begin{align}
\begin{split}
   \mathrm{Prob}(0) &= \frac{1}{4}\braket{\psi_i| V_k^\dagger V_k |\psi_i} + \frac{1}{4}+ \frac{1}{4}\braket{\psi_i| V_k^\dagger |\psi_f}+ \frac{1}{4}\braket{\psi_f| V_k |\psi_i}\,,\\
   \mathrm{Prob}(1) &= \frac{1}{4}\braket{\psi_i| V_k^\dagger V_k |\psi_i} + \frac{1}{4}- \frac{1}{4}\braket{\psi_i| V_k^\dagger |\psi_f}- \frac{1}{4}\braket{\psi_f| V_k |\psi_i}\,,   
\end{split}\end{align}
and finally
\begin{align}
\begin{split}
   \mathrm{Prob}(0)- \mathrm{Prob}(1) &= \frac{1}{2}\braket{\psi_i| V_k^\dagger |\psi_f} + \frac{1}{2}\braket{\psi_f| V_k |\psi_i} 
   \equiv \mathrm{Re}[\braket{\psi_f| V_k |\psi_i}]\,.  
\end{split}\end{align}
The same results can be recovered following the circuit for the imaginary part.

Lastly, we discuss the complexity scaling of the proposed approach. As mentioned for the encoding, for an $N_q$-qubit system, our Hamiltonian would only consist of $\mathcal{O}(N_q^2)$ two-qubit Pauli terms.
Consequently, simulating the time evolution operators $U_i$ and $U_f^\dagger$ via a Trotter-Suzuki decomposition requires $\mathcal{O}(N_q^2)$ two-qubit gates per Trotter step. To reach a total simulation time $t$ (equal to $L$ in this case) within an algorithmic Trotter error tolerance $\epsilon_{\text{trotter}}$, the evolution is discretized into several $N_s$ trotter steps per circuit execution.
Note the controlled unitaries share the same asymptotic as the uncontrolled one, with a larger constant prefactor~\cite{Barenco1995}. 
Since the controlled-$V_k$ operation adds only an $\mathcal{O}(1)$ gate overhead to the circuit, the asymptotic gate complexity for the circuit in Fig.~\ref{fig:hadamard-circuit} to evaluate each $V_k$ scales as $\mathcal{O}( N_q^2 t^{1+1/p}/\epsilon_{\text{trotter}}^{1/p})$ for an $p$-order trotterization. 

\section{Numerical results}\label{sec:results}

\begin{figure}[htp!]
    \centering
    \begin{subfigure} {.99\textwidth}
        \centering
        \includegraphics[width=0.49\textwidth]{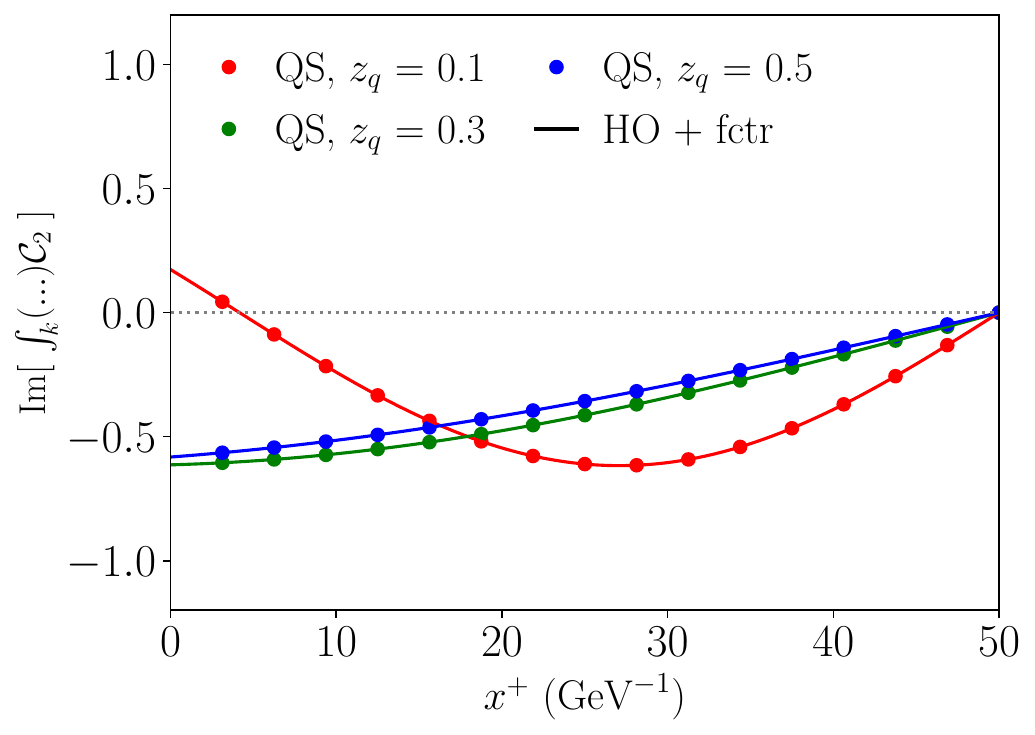}
        \includegraphics[width=0.49\textwidth]{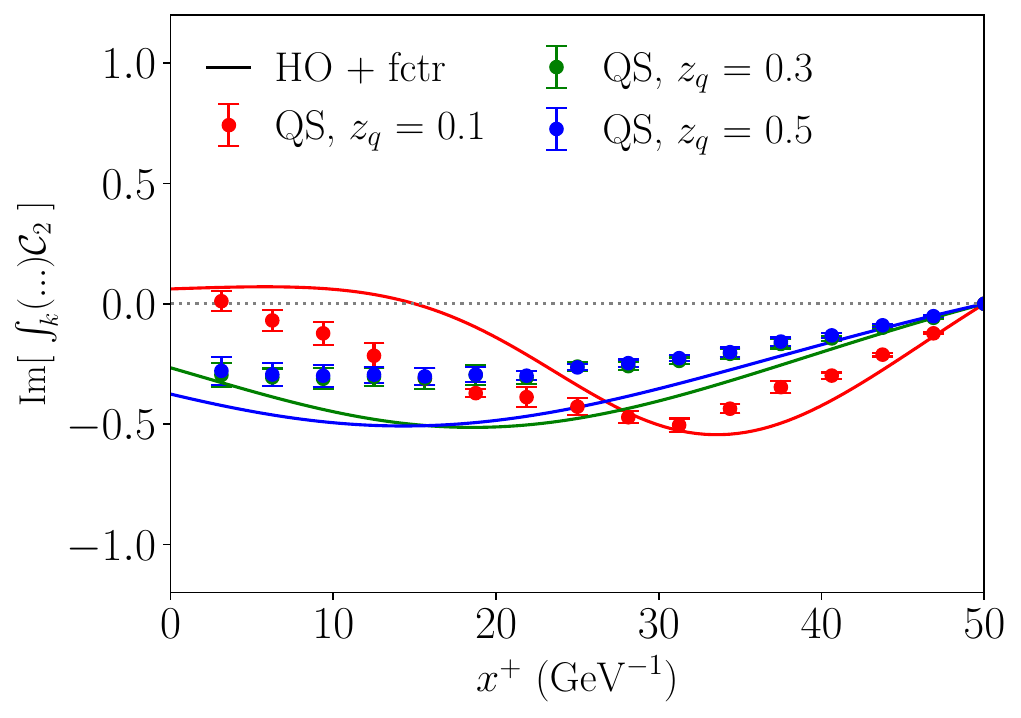}
        \caption{momentum-integrated $\mathcal{C}_2$ }
    \end{subfigure}
    \begin{subfigure} {.99\textwidth}
        \centering
        \includegraphics[width=0.49\textwidth]{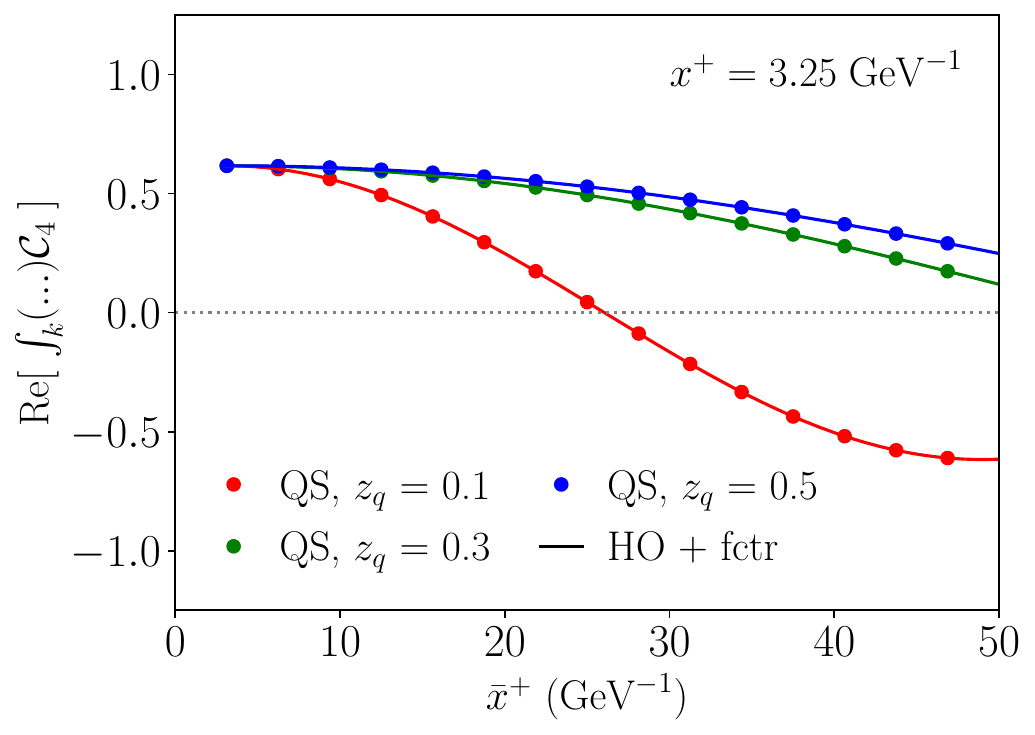}
        \includegraphics[width=0.49\textwidth]{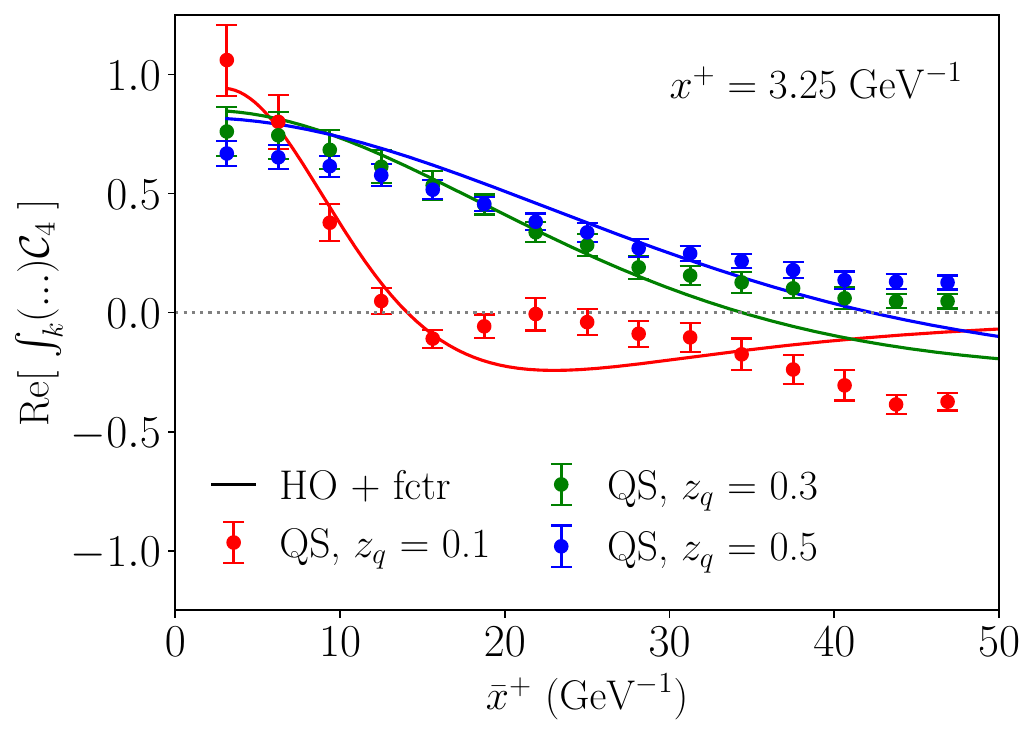}
        \caption{momentum-integrated $\mathcal{C}_4$}
    \end{subfigure}
    \begin{subfigure} {.99\textwidth}
        \centering
        \includegraphics[width=0.49\textwidth]{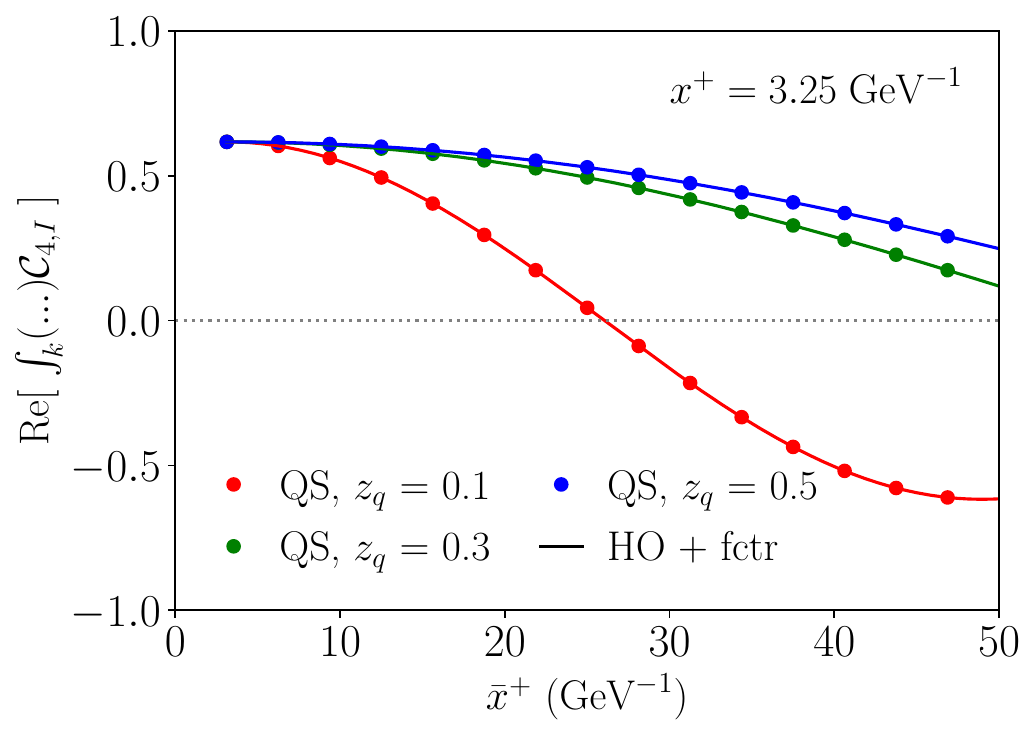}
        \includegraphics[width=0.49\textwidth]{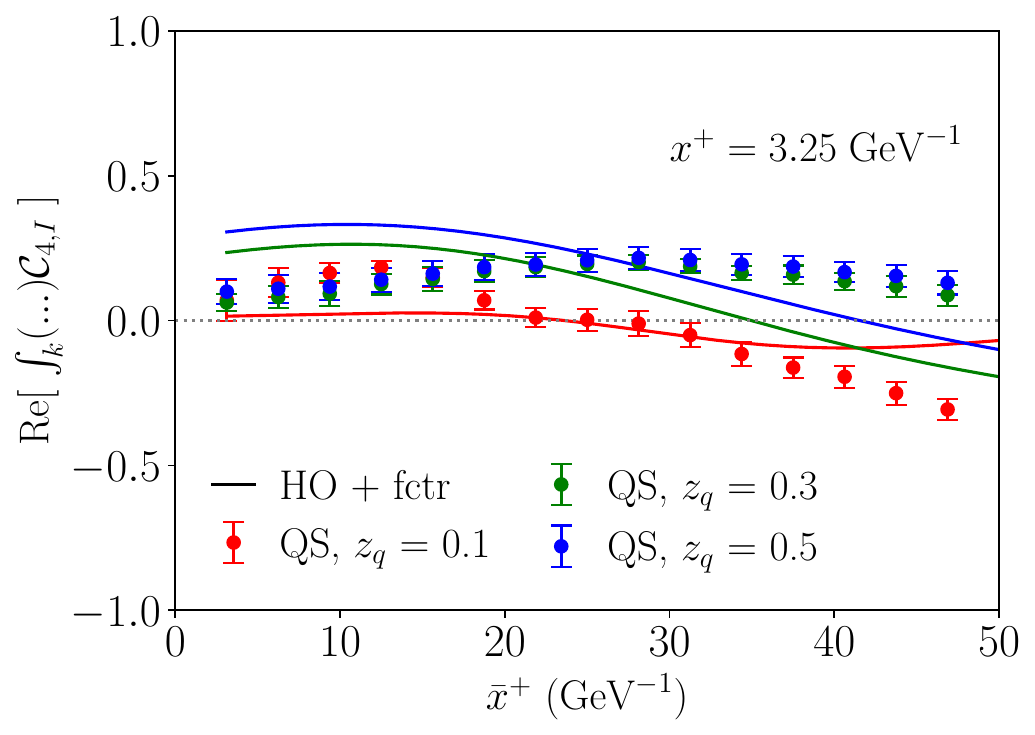}
        \caption{momentum-integrated $\mathcal{C}_{4,I}$}
    \end{subfigure}
    \caption{
    \textbf{Simulation results for the momentum-integrated matrix elements.} The different panels label the three studied correlators: (a) $\mathcal{C}_2$, (b) $\mathcal{C}_4$, and (c) $\mathcal{C}_{4,I}$.
    The left panels show the vacuum results, while the right panels show the corresponding results in the medium with $\hat{q}=\hat q_2$.
    Results from the quantum simulation are shown in dots, with different colors corresponding to different values of $z$, while the analytical results are shown as solid lines.
    \label{fig:result-matrix-elements}
    }
\end{figure}

In this section, we present the numerical calculation using the quantum circuit approach introduced above, via classical simulation methods. Carrying out the smallest nontrivial calculation with $N_c=2$ colors using the quantum circuit would require $N_q = 8N_cN_\perp^2> 64$ qubits, well beyond the resources available on noise-free quantum emulators. For this reason, we instead present quantum simulation (QS) results using exact diagonalization, storing the full wave function explicitly.
We compare these exact results with the corresponding analytical estimates  discussed in \cref{sec:mapping_to_QIS}, noting that the latter are obtained under additional simplifications: the HO approximation, the large-$N_c$ limit and the factorization of the quadrupole correlator under a specific kinematic limit. Where available, we also include results obtained under a semi-classical approximation~\cite{Dominguez:2019ges}.

We use the following parameters throughout this section for the simulations, unless otherwise specified. The transverse lattice size is $N_\perp=2$, $L_\perp=4\; \mathrm{GeV}^{-1}$.
The associated transverse momentum space thereby spans from $-N_\perp\pi/L_\perp=-1.57$ GeV to $(N_\perp-1)\pi/L_\perp=1.57$ GeV with a resolution of $d_p=0.785$ GeV.
The longitudinal momentum of the dipole system is $p^+=50$ GeV.
The medium extension is $L=50 \;\mathrm{GeV}^{-1}$ and the IR regulator $m_g=0.8$ GeV. 
The time evolution is performed with $N_s=16$ steps.
For the results in the medium, we average over ten independent medium configurations and take the mean value, with the standard error shown as the uncertainty bar/band. 
Note that both simulation and analytical results use the same particle density $n_0$ as input, for the former this is a direct input to the correlation function in \cref{eq:MV_color_charge}, while for the latter it is reparameterized as $\hat q_0$ according to \cref{eq:qhat0}. In all comparisons, the three representative values for $\hat q$ according to \cref{eq:qhat} are:
$\hat{q}_1=0.0057 \;\mathrm{GeV}^3$, $\hat{q}_2=0.0169 \;\mathrm{GeV}^3$, and  $\hat{q}_3=0.0300 \;\mathrm{GeV}^3$. These values are related to $\hat q_0$ used in the analytical calculation as $\hat{q} \approx 0.785\,\hat{q}_0$.
We note that the values of $\hat{q}$ are relatively small compared to typical values encountered in QGP calculations, owing to the limited lattice size.

\subsection{Matrix element of the correlator}

We first examine the individual terms that contribute to the modification factors $F_{\rm med}$ and $F_{\rm med,g }$. Specifically, we calculate the momentum-integrated matrix elements $\mathcal{C}_2$, $\mathcal{C}_4$, and $\mathcal{C}_{4,I}$ as functions of time before implementing the time integrations. 
Although these quantities do not carry a direct physical interpretation, they are the primary outputs of our quantum simulation algorithm and serve as the building blocks for constructing the modification factors.

In \cref{fig:result-matrix-elements}, we present the results for the momentum-integrated matrix elements, $\mathcal{C}_2$ and $\mathcal{C}_4$, and $\mathcal{C}_{4,I}$ (from top to bottom rows), as defined in the LHS of \cref{eq:k_int_C2_C4,eq:k_int_C4_C4I}.
The left panels show the vacuum results, while the right panels correspond to the medium case with $\hat{q}=\hat q_2$.
We fix the relative momentum squared to $\p^2=0.62\; \mathrm{GeV}^2$, such that the particle momenta lie near the center of the lattice and are therefore less affected by boundary artifacts. 
Results from the quantum simulation (labeled as QS in the plot legends) are shown as dots, with different colors corresponding to different values of $z$, while the analytical results, obtained according to the RHS of \cref{eq:k_int_C2_C4,eq:k_int_C4_C4I}, are shown as solid lines.

For the vacuum case, the simulation results for all three quantities agree very well with the analytical expectations, providing a benchmark for our simulation framework.
In contrast, in the presence of the medium, the simulation results deviate from the analytical estimates, as expected. 

\subsection{Modification factor for dipole formation}\label{sec:results_dijet}

\begin{figure}[htp!]
    \centering
    \begin{subfigure}{0.49\textwidth}
        \centering
        \includegraphics[width=0.99\textwidth]{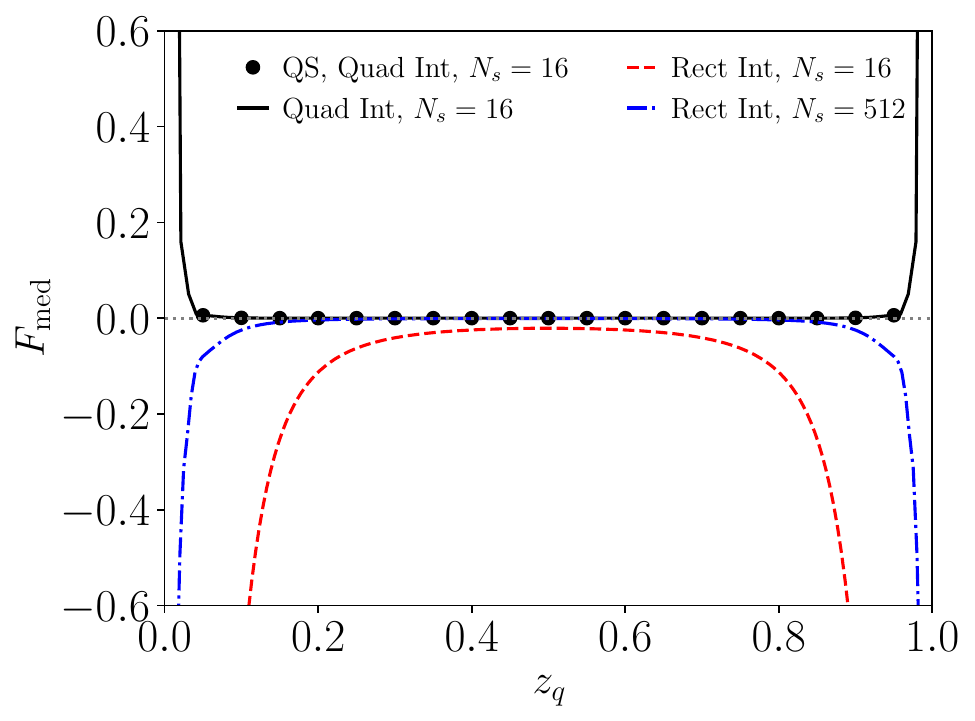}
        \subcaption{$\p^2=0.62\; \mathrm{GeV}^2$}
    \end{subfigure}
    \begin{subfigure}{0.49\textwidth}
        \centering
        \includegraphics[width=0.99\textwidth]{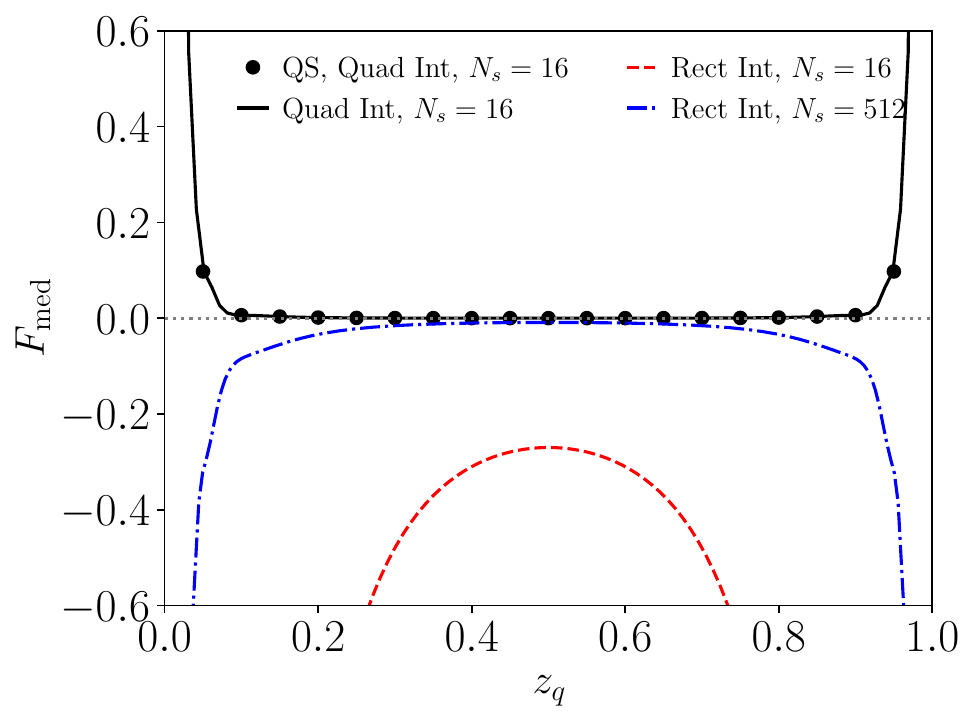}
        \subcaption{$\p^2=1.23\; \mathrm{GeV}^2$}
    \end{subfigure}
    \caption{\textbf{Simulation results of $F_{\mathrm{med}}$ in vacuum.} The QS results with $N_s = 16$ steps using Gauss-Legendre quadrature method after applying cubic interpolation on the integrand are compared with HO + fctr analytical results at various $N_s$ using the same quadrature method and with rectangular integration. The true $F_\mathrm{med} =0$ at the continuum. 
    }
    \label{fig:result-vac}
\end{figure}

Having examined the contributing matrix elements, we now turn to the modification factor for the quark-antiquark dipole formation.
To benchmark the time integration method, in \cref{fig:result-vac} we perform a detailed investigation of the deviation from the exact result in the vacuum limit, i.e., $F_{\rm med} = 0$. 
The black dots correspond to the simulation results obtained with Gauss-Legendre quadrature after cubic interpolation of the time-dependent integrands evaluated with $N_s = 16$ time steps,
while the analytical results using the same quadrature method are shown as black solid lines. 
Importantly, the QS results for both values of $p_\perp^2 \in {0.62,, 1.23}$ GeV$^2$ agree with the exact result over most of the $z_q$ range, consistent with what was observed for the matrix elements in \cref{fig:result-matrix-elements}, with sizable deviations only in the limits $z_q \to 0, 1$. 
Since the integrand is evaluated with a finite number of time steps, achieving the exact result in these limits is challenging, as the limited number of time evaluations does not allow the interpolation to fully capture the resulting oscillatory behavior. 
Nevertheless, these numerical deviations are similar between the QS and analytical results, as confirmed by the superposition of the black dots and line. 
Furthermore, for comparison, analytical results using rectangular integration (Rect Int) at $N_s = 16$ and $512$ are also provided, exhibiting strong discretization artifacts near the edges of the $z_q$ distribution and poor convergence even at $N_s = 512$, in contrast to the quadrature method (Quad Int) at $N_s = 16$. 
Since the computation of $F_\mathrm{med}$ and $F_\mathrm{med,g}$ requires $\mathcal{O}(N_s^2)$ evaluations of the matrix elements in QS, efficient time integration is essential, and we therefore adopt the quadrature method for all subsequent calculations. 

\begin{figure}[htp!]
    \centering
    \begin{subfigure}{0.49\textwidth}
        \centering
        \includegraphics[width=0.99\textwidth]{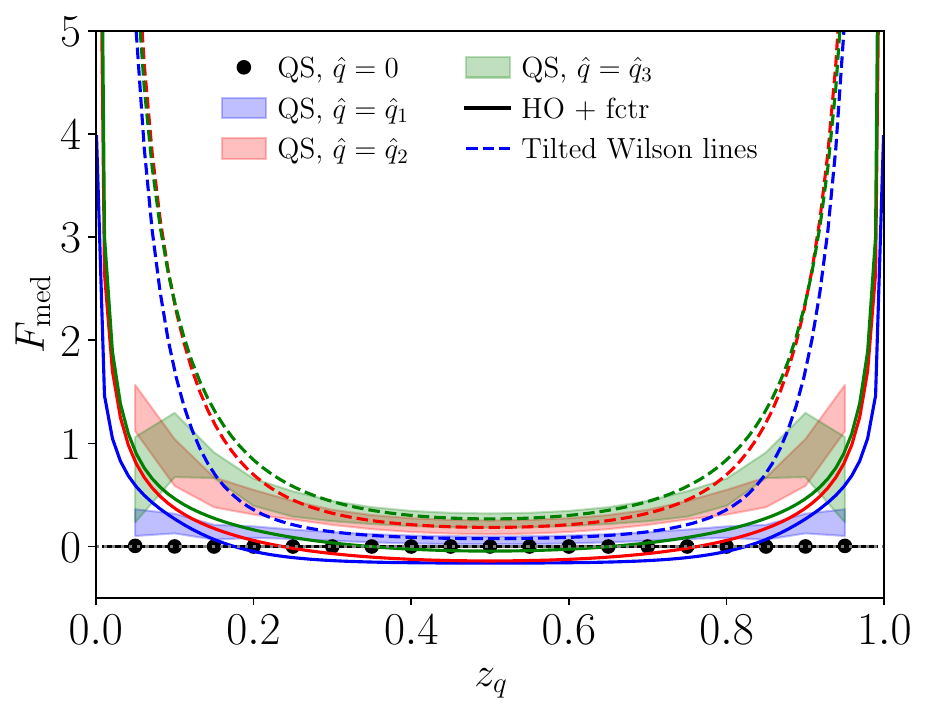}
        \subcaption{$p^2=0.62\; \mathrm{GeV}^2$\label{fig:Fmed_a}}
    \end{subfigure}
    \begin{subfigure}{0.49\textwidth}
        \centering
        \includegraphics[width=0.99\textwidth]{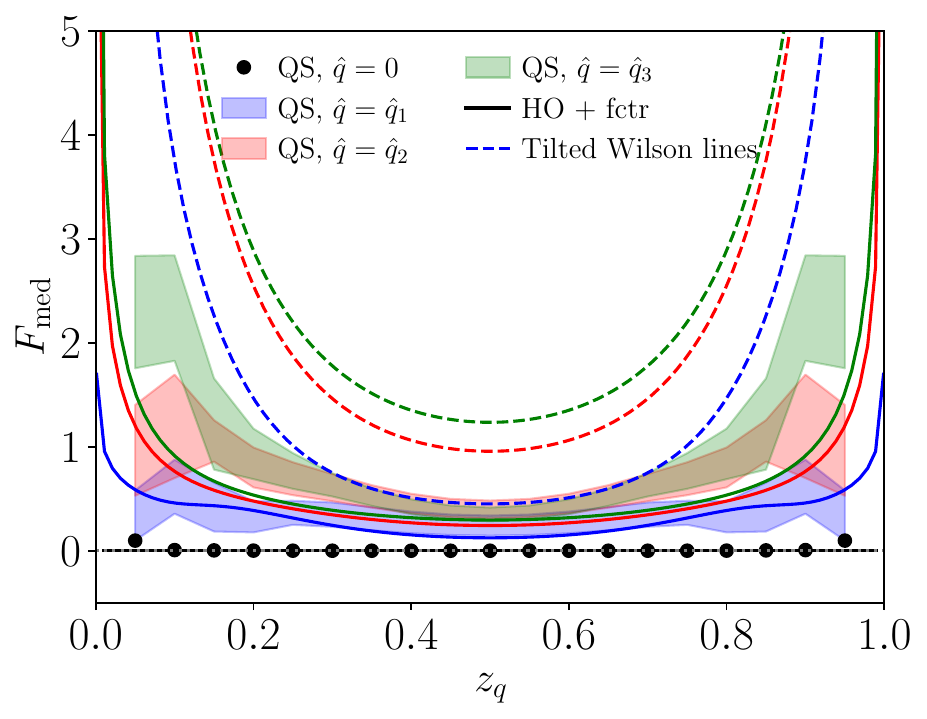}
        \subcaption{$p^2=1.23\; \mathrm{GeV}^2$\label{fig:Fmed_b}}
    \end{subfigure}
    \caption{\textbf{Simulation results of the modification factor $F_{\mathrm{med}}$ for various $\hat{q}$}. Two values of relative momentum squared are provided with $\hat{q}_1=0.0057 \;\mathrm{GeV}^3$, $\hat{q}_2=0.0169 \;\mathrm{GeV}^3$, and  $\hat{q}_3=0.0300 \;\mathrm{GeV}^3$. 
    \label{fig:result-Fmed}
    }
\end{figure}

In \cref{fig:result-Fmed} we present
the modification factor $F_{\rm med}$ defined in \cref{eq:Fmed} in the vacuum limit ($\hat q = 0$) and also in the presence of a medium. 
We choose two values of the relative momentum squared, $\p^2=0.62 \GeV^2$ and $1.23 \GeV^2$, shown in \cref{fig:Fmed_a} and \cref{fig:Fmed_b}, respectively. 
The simulation results in vacuum are shown as black dots, while the results in the medium are shown as colored bands corresponding to different values of $\hat q$. The band widths indicate the uncertainty in terms of standard errors (SE = $\sigma/\sqrt{N_\mathrm{event}}$) arising from sampling from $N_\mathrm{event}=10$ medium configurations. 
The analytical results labeled ``HO+fctr'', obtained from \cref{eq:Fmed_ana}, are shown as solid lines, with colors matching the corresponding QS results.
The dashed lines correspond to the tilted Wilson line (semi-classical) approximation, where all particles are assumed to be sufficiently energetic (namely $p_i^+ \gg L^{-1}$) such that they follow classical paths and the medium mainly causes color rotation along their trajectories, see e.g.~\cite{Dominguez:2019ges} for details on this kinematical approximation. 

In the presence of the medium, the simulation results show sizable difference from the analytical estimates.
In the case of smaller $\p^2$, as shown in \cref{fig:Fmed_a}, when $\hat q$ increases, the simulation results indicate a stronger positive medium modification in the intermediate $z_q$ regime, whereas the HO + fctr analytical results predict a weaker $\hat q$ dependence, which for intermediate $z_q$ actually results in a smaller suppression of the cross-section with increasing $\hat q$. Note that the divergence for small $z_q$ is expected analytically. Physically, this is a consequence of an increasingly soft dipole pair ($z_q\ll1$) being forced to carry a finite transverse momentum $\p^2$. This kinematic setup is naturally not captured by the collinear limit underlying both calculations.  
The semi-classical results, labeled as ``Tilted Wilson lines", are in better agreement with the QS results for $z_q \sim 0.5$, although severely overestimating $F_{\rm med}$ for $z_q\lesssim 0.1$. This is expected, since only for balanced enough splittings are both $q$ and $\bar q$ sufficiently energetic to be in the semi-classical regime.
In contrast, in the larger $\p^2$ case, as shown in \cref{fig:Fmed_b}, all three results yield positive medium modification in the intermediate $z_q$ regime, with the semi-classical results overestimating the magnitude of $F_{\rm med}$ and its dependence on $\hat q$ with respect to the remaining two sets of results - QS and HO+fctr. These two roughly agree on the sensitivity to the value of $\hat q$, with the increase from $\hat q_2$ to $\hat q_3$ being quite modest, and even agree on the strength of the modification for the smallest value $\hat q=\hat q_1$. For larger $\hat q$, the HO\,+\,fctr result underestimates $F_{\rm med}$. 
As previously, there is divergent behaviour in the limits $z_q \to 0,\,1$, which corresponds to large splitting angles outside of the present calculation's validity. 
Finally, since the analytical results rely on additional simplifications, the observed discrepancy reflects the limitations of these approximations and highlights the more general nature of the quantum simulation.

\subsection{Modification factor for color decoherence of QCD antennas}\label{sec:results_antena}

\begin{figure}[t]
    \centering
    \begin{subfigure}{0.49\textwidth}
        \centering
        \includegraphics[width=0.99\textwidth]{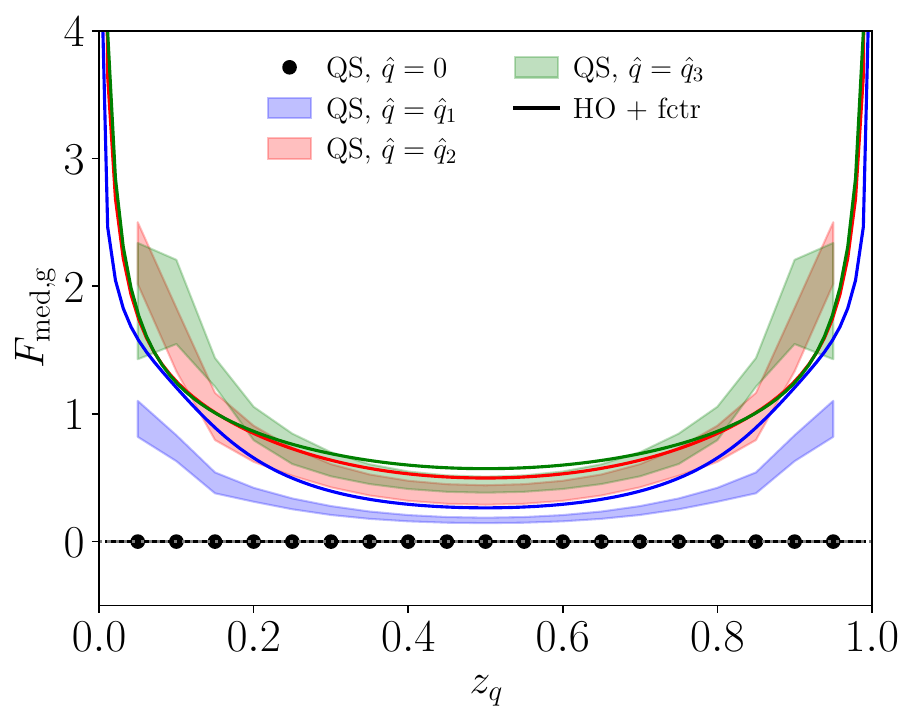}
        \subcaption{$\p^2=0.62\; \mathrm{GeV}^2$\label{fig:Fmedg_a}}
    \end{subfigure}
    \begin{subfigure}{0.49\textwidth}
        \centering
        \includegraphics[width=0.99\textwidth]{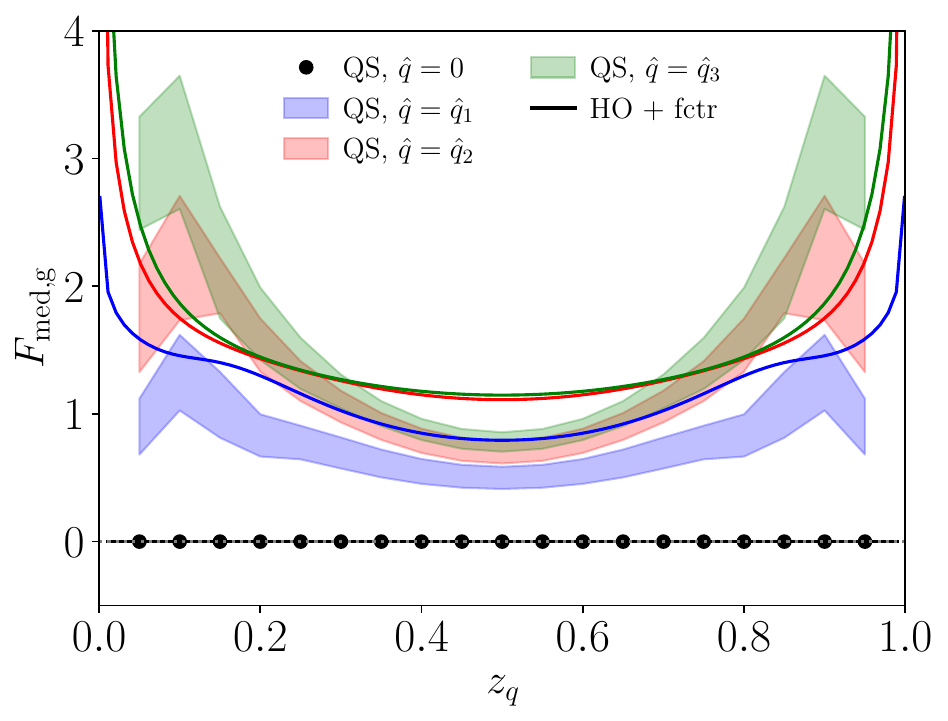}
        \subcaption{$\p^2=1.23\; \mathrm{GeV}^2$\label{fig:Fmedg_b}}
    \end{subfigure}
    \caption{
    \textbf{Results of the modification factor $F_{\mathrm{med,g}}$ for various $\hat{q}$.} Two values of relative momentum squared are provided with $\hat{q}_1=0.0057 \;\mathrm{GeV}^3$, $\hat{q}_2=0.0169 \;\mathrm{GeV}^3$, and  $\hat{q}_3=0.0300 \;\mathrm{GeV}^3$.
    \label{fig:result-Fmedg}
    }
\end{figure}

Finally, we consider the modification factor for the soft gluon emission off a singlet $q\bar q$ dipole.
We present in \cref{fig:result-Fmedg} the modification factor $F_{\rm med,g}$ defined in \cref{eq:F_med_anti_ang}. 
As in the case of $F_{\rm med}$, we choose two values of the relative momentum squared, $\p^2=0.62\,\GeV^2$ and $1.23\,\GeV^2$, shown in
\cref{fig:Fmedg_a} and \cref{fig:Fmedg_b}, respectively. 
The HO + fctr analytical results are obtained according to \cref{eq:F_med_anti_ang_ana}.
The plotting conventions are the same as in \cref{fig:result-Fmed}. As with $F_{\rm med}$, the simulation results for $F_{\rm med,g}$ agree very well with the analytical estimates in the vacuum case, but show notable deviations for the absolute values in the medium, again reflecting the limitations of the analytical approximations used for comparison. Nevertheless, the sensitivity to $\hat q$ is roughly similar between the two sets of results.
Lastly, we emphasize that, given the small lattice size considered here, the QS results should not be interpreted as quantitatively superior to the analytic estimates in the medium case. A full quantitative assessment would require significantly larger lattice size, which lies beyond the scope of the present work.

\section{Conclusion}\label{sec:conclusion}
In this work we have developed a framework to compute multi-particle processes in high-energy nuclear environments using quantum simulation methods. By mapping partonic amplitudes onto unitary time evolution in a light-front Hamiltonian picture, we formulated the interaction of energetic probes with QCD matter directly at the amplitude level. This approach naturally captures the full color structure of the process and avoids the need for ensemble-level approximations or further simplifications of the color structure, as is commonly employed in other treatments.

We benchmarked the framework in two representative cases: dipole formation and the radiation pattern of QCD antennas in a medium. In both situations, we expressed the relevant observables in terms of matrix elements --- $\cC_2$, $\cC_4$, and $\cC_{4,I}$ --- that admit a direct implementation as quantum circuits. The comparison of the analytic curves with the quantum simulation results allows to gauge the magnitude of the approximations used in the analytic approach, and reveals the importance of computing the exact matrix elements. Nonetheless, one should keep in mind that the substantial differences found between both results might also result from truncation and lattice artifacts not fully quantified due to the small system sizes considered.

Beyond these benchmarks, the present formulation provides a systematic and extensible pathway to study increasingly more complex multi-particle processes in realistic QCD backgrounds. In particular, the ability to evaluate correlators without relying on Gaussian statistics, large-$N_c$ expansion, or factorization assumptions opens the door to systematically testing some of these approximations further in more complex final states. We note that since the simulations are performed at the amplitude level, this approach may be better suited for exploring jet evolution in non-trivial backgrounds~\cite{Altenburger:2025iqa,Avramescu:2025lhr,Avramescu:2026fgv, Carrington:2026qlg,Barata:2024xwy,Ipp:2020nfu}, which are harder to capture working directly with medium-averaged quantities.

\paragraph*{Additional note:} While completing this manuscript, we became aware of related work~\cite{Alba:future}, where a complete computation of the leading order in-medium QCD splitting kernels is performed at finite number of colors and using the full form of the in-medium scattering rate, similarly to what is done in this work for the quark channel. The main difference between these two approaches lies that while our calculation is set at amplitude level, i.e. using the gauge field $\mathcal{A}$ directly, the computation reported in~\cite{Alba:future} is performed after averaging over field configurations, i.e. the computation is carried out at the level of $\langle \mathcal{A} \mathcal{A}\rangle$. One should note that the computation in~\cite{Alba:future} does not require the construction of a quantum circuit and thus, due to the current limitation on quantum simulation resources, the results reported there go beyond our current numerical capabilities. 

\acknowledgments
We thank  Marco Leitão and Alba Soto-Ontoso for clarifications on their work~\cite{Alba:future} and for allowing us to cross-check the numerical results. ML, WQ, CS and JMS are supported by European Research Council under project ERC-2018-ADG-835105 YoctoLHC; by Maria de Maeztu excellence unit grant CEX2023-001318-M and project PID2023-152762NB-I00 funded by MICIU/AEI/10.13039/501100011033; by ERDF/EU; and by Xunta de Galicia (CIGUS Network of Research Centres). 
ML and WQ acknowledge the support of Xunta de Galicia under the ED431F 2023/10 project.
WQ is also supported by  the Marie Sklodowska-Curie Actions Postdoctoral Fellowships under Grant No. 101109293.
JMS is also supported by OE - Portugal, Fundação para a
Ciência e Tecnologia (FCT) under contract PRT/BD/152262/2021, and by ERDF (grant PID2022-139466NB-C21) and Consejería de Universidad, Investigación e Innovación, Gobierno de España and Unión Europea – NextGenerationEU under grant AST22 6.5.

\bibliographystyle{JHEP-2modlong.bst}
\bibliography{refs.bib}

\end{document}